\documentclass[reprint,superscriptaddress,amsmath,amssymb,aps,longbibliography]{revtex4-1}

\usepackage[colorlinks]{hyperref}
\usepackage{graphicx}
\usepackage{dcolumn}
\usepackage{bm}
\usepackage{booktabs}
\usepackage{array}
\usepackage{color}

\def\avg#1{\mathinner{\langle{#1}\rangle}}
\newcommand{\eq}[1]{Eq.~\hyperref[eq:#1]{(\ref*{eq:#1})}}
\renewcommand{\sec}[1]{\hyperref[sec:#1]{Section~\ref*{sec:#1}}}
\newcommand{\app}[1]{\hyperref[app:#1]{Appendix~\ref*{app:#1}}}
\newcommand{\tab}[1]{\hyperref[tab:#1]{Table~\ref*{tab:#1}}}
\newcommand{\fig}[1]{\hyperref[fig:#1]{Figure~\ref*{fig:#1}}}
\newcommand{\figa}[2]{\hyperref[fig:#1]{Figure~\ref*{fig:#1}#2}}
\newcommand{\figx}[2]{\hyperref[fig:#1]{Figure~\ref*{fig:#1}(#2)}}
\newcommand{\thm}[1]{\hyperref[thm:#1]{Theorem~\ref*{thm:#1}}}
\newcommand{\lem}[1]{\hyperref[lem:#1]{Lemma~\ref*{lem:#1}}}
\newcommand{\cor}[1]{\hyperref[cor:#1]{Corollary~\ref*{cor:#1}}}
\newcommand{\defn}[1]{\hyperref[def:#1]{Definition~\ref*{def:#1}}}
\newcommand{\alg}[1]{\hyperref[alg:#1]{Algorithm~\ref*{alg:#1}}}

\begin{document}

\title{Efficient and Noise Resilient Measurements for Quantum Chemistry\texorpdfstring{\\}{} on Near-Term Quantum Computers}

\author{William J. Huggins}
\email{corresponding author: wjhuggins@gmail.com}
\affiliation{Google Research, Venice, CA 90291}
\affiliation{Department of Chemistry and Berkeley Quantum Information and Computation Center, University of California, Berkeley, CA 94720}
\author{Jarrod R. McClean}
\affiliation{Google Research, Venice, CA 90291}
\author{Nicholas C. Rubin}
\affiliation{Google Research, Venice, CA 90291}
\author{Zhang Jiang}
\affiliation{Google Research, Venice, CA 90291}
\author{Nathan Wiebe}
\affiliation{Google Research, Venice, CA 90291}
\affiliation{Paciﬁc Northwest National Laboratory, Richland WA 99354}
\affiliation{Department of Physics, University of Washington, Seattle, WA 98105}
\author{K. Birgitta Whaley}
\affiliation{Department of Chemistry and Berkeley Quantum Information and Computation Center, University of California, Berkeley, CA 94720}
\author{Ryan Babbush}
\email{corresponding author: ryanbabbush@gmail.com}
\affiliation{Google Research, Venice, CA 90291}

\date{\today}

\begin{abstract} Variational algorithms are a promising paradigm for utilizing
near-term quantum devices for modeling electronic states of molecular systems. However, previous
bounds on the measurement time required have suggested that the application of
these techniques to larger molecules might be infeasible. We present a
measurement strategy based on a low rank factorization of the two-electron
integral tensor. Our approach provides a cubic reduction in term groupings over
prior state-of-the-art and enables measurement times three orders of magnitude
smaller than those suggested by commonly referenced bounds for the largest
systems we consider. Although our technique requires execution of a linear-depth
circuit prior to measurement, this is compensated for by eliminating challenges
associated with sampling non-local Jordan-Wigner transformed operators in the
presence of measurement error, while enabling a powerful form of error
mitigation based on efficient postselection. We numerically characterize these
benefits with noisy quantum circuit simulations for ground state energies of strongly correlated electronic systems.
\end{abstract}

\maketitle

\section{Introduction}

Given the recent progress in quantum computing hardware, it is natural to ask
where the first demonstration of a quantum advantage for a practical problem
will occur. Since the first experimental demonstration by Peruzzo et
al.~\cite{peruzzo2014variational}, the variational quantum eigensolver (VQE)
framework has offered a promising path towards utilizing small and noisy quantum
devices for simulating quantum chemistry. The essence of the VQE approach is the
use of the quantum device as a co-processor which prepares a parameterized
quantum wavefunction and measures the expectation value of observables. In
conjunction with a classical optimization algorithm, it is possible to then
minimize the expectation value of the Hamiltonian as a function of the
parameters, arriving at approximations for the wavefunction, energy, and other
properties of the ground state~\cite{peruzzo2014variational, mcclean2016theory,
wecker2015progress,OMalley2016, kandala2017hardware, lee2018generalized,
parrish2019quantum,o2019calculating}. A growing body of work attempting to
understand and ameliorate the challenges associated with using VQE to target
non-trivial systems has emerged in recent years~\cite{bonet2018low,
mcclean2019decoding, mcardle2019error, temme2017error, sagastizabal2019error,
mcclean2017hybrid, otten2019accounting, o2019generalized, 1907.03358,
1907.07859, 1907.09386, 1907.09040,izmaylov2019revising, Gokhale2019-cv}. In
this article we address the challenge posed by the large number of
circuit repetitions needed to perform accurate measurements and propose a new
scheme that dramatically reduces this cost. Additionally, we explain how our
approach to measurement has reduced sensitivity to readout errors and also
enables a powerful form of error mitigation.

Within VQE, expectation values are typically estimated by Hamiltonian averaging.
Under this approach, the Hamiltonian is decomposed into a sum of operators that
are tensor products of single-qubit Pauli operators, commonly referred to as
Pauli strings. The expectation values of the Pauli strings are determined
independently by repeated measurement. When measurements are distributed
optimally between the Pauli strings $P_\ell$, the total number of measurements
$M$ is upper bounded by
\begin{equation} M \leq \left(\frac{\sum_\ell
\left|\omega_\ell\right|}{\epsilon}\right)^2, \quad \textrm{where} \quad H =
\sum_\ell \omega_\ell P_\ell
\label{eq:L1_bound}
\end{equation} is the Hamiltonian whose expectation value we estimate as
$\sum_{\ell} \omega_\ell \avg{P_\ell}$, the $\omega_\ell$ are scalars, and
$\epsilon$ is the target precision
\cite{wecker2015progress,rubin2018application}. Prior work assessing the
viability of VQE has used bounds of this form and concluded that chemistry
applications require ``a number of measurements which is astronomically large''
(quoting from Ref.~\citenum{wecker2015progress}).

\begin{table*}[t]
\begin{tabular}{l|c|c|c|c|c|c|c}

Ref. & Partitioning Method & Circuit Description & \# of Partitions & Gate Count
& Depth & Connectivity & Diagonal \\ \hline\hline \cite{mcclean2016theory} &
commuting Pauli heuristic & - & $O(N^4)$ & - & - & - & -\\
\cite{kandala2017hardware} & compatible Pauli heuristic & single rotations &
$O(N^4)$ & $N$ & 1 & any & no\\
\cite{rubin2018application} & $n$-representability constraints & single
rotations & $O(N^4)$ & $N$ & 1 & any & no\\
\cite{izmaylov2019revising} & mean-field partitioning & fast feed-forward &
$O(N^4)$ & $O(N)$ & $O(N)$ & full & no\\
\cite{1907.03358} & compatible Pauli clique cov. & single rotations & $O(N^4)$ &
$N$ & $1$ & any & no\\
  \cite{o2019generalized} & counting argument & swap networks & $O(N^3)$ &
  $O(N^2)$ & $O(N)$ & linear & no\\
\cite{1907.07859} & commuting Pauli graph color. & stabilizer formalism &
$O(N^3)$ & - & - & full & no\\
\cite{1907.09040} & anticommuting Pauli clique cov. & Pauli evolutions &
$O(N^3)$ & $O(N^2 \log(N))$ & - & full & no\\
 \cite{1907.09386} & commuting Pauli clique cover & symplectic subspaces &
 $O(N^3)$ & $O(N^2 / \log N)$ & - & full & no\\
\cite{Gokhale2019-cv} & commuting Pauli clique cover & stabilizer formalism &
$O(N^3)$ & $O(N^2)$ & - & full & no\\
 here & integral tensor factorization & Givens rotations & $O(N)$ & $N^2 / 4$ &
 $N / 2$ & linear & yes\\
\hline
\end{tabular}
\label{tab:comparison_other_works}
\caption{A history of ideas reducing the measurements required for estimating
the energy of arbitrary basis chemistry Hamiltonians with the variational
quantum eigensolver. Here $N$ represents the number of spin-orbitals in the
basis. Gate counts and depths are given in terms of arbitrary 1- or 2-qubit
gates restricted to the geometry of 2-qubit gates specified in the connectivity
column. What we mean by ``compatible'' Pauli groupings is that the terms can be
measured at the same time with only single qubit rotations prior to measurement.
We report whether terms are measured in a diagonal representation as this is
important for enabling strategies of error-mitigation by postselection. The
number of partitions refers to the number of unique term groupings which can
each be measured with a single circuit - thus, this reflects the number of
unique circuits required to generate at least one sample of each term in the
Hamiltonian. However, we caution that one cannot infer the total number of
measurements required from the number of partitions, and often this metric is
highly misleading. The overall number of measurements required is also
critically determined by the variance of the estimator of the energy. As
explained in the first entry of this table, when terms are measured
simultaneously one must also consider the covariance of those terms. In some
cases, a grouping strategy can decrease the number of partitions but increase
the total number of measurements required by grouping terms with positive
covariances. Alternatively, strategies such as the third entry in this table
actually increase the number of partitions while reducing the number of
measurements required overall by lowering the variance of the estimator.}
\end{table*}

Several recent proposals attempt to address this obstacle by developing more
sophisticated strategies for partitioning the Hamiltonian into sets of
simultaneously measurable operators~\cite{izmaylov2019revising,
  o2019generalized, 1907.03358, 1907.07859, 1907.09386,
  1907.09040}. We summarize their key findings in \tab{comparison_other_works}.
This work has a similar aim, but we take an approach rooted in a decomposition
of the two-electron integral tensor rather than focusing on properties of Pauli
strings. We quantify the performance of our proposal by numerically simulating
the variances of our term groupings to more accurately determine the number of
circuit repetitions required for accurate measurement of the ground state
energy. This contrasts with the analysis in other recent papers that have
instead focused on using the number of separate terms which must be measured as
a proxy for this quantity. By that metric, our approach requires a number of
term groupings that is linear in the number of qubits - a quartic improvement
over the naive strategy and a cubic improvement relative to these recent papers.
However, we argue that the number of distinct term groupings alone is not
generally predictive of the total number of circuit repetitions required,
because it does not consider how the covariances of the different terms in these
groupings can collude to either reduce or increase the overall variance. We will
show below that our approach benefits from having these covariances conspire in
our favor; for the systems considered here our approach gives up to three orders
of magnitude reduction in the total number of measurements, while also providing
an empirically observed asymptotic improvement.

Although there are a variety of approaches to simulating indistinguishable
fermions with distinguishable qubits~\cite{setia2018bravyi,
  jiang2018majorana,bravyi2002fermionic}, the Jordan-Wigner transformation is
the most widely used. 
This is due to its simplicity and to the fact that it allows for the explicit
construction of a number of useful circuit primitives not available under more
sophisticated encodings. 
These include the Givens rotation network that exactly implements a change of
single-particle
basis~\cite{babbush2018low,o2019generalized,kivlichan2018quantum,motta2018low}.
A disadvantage of using the Jordan-Wigner transformation is the fact that it
maps operators acting on a constant number of fermionic modes to qubit operators
with support on up to all \(N\) qubits. 
In the context of measurement, the impact of this non-locality can be seen by
considering a simple model of readout error such as a symmetric bitflip channel. 
Under this model, a Pauli string with support on \(N\) qubits has \(N\)
opportunities for an error that reverses the sign of the measured value, leading
to estimates of expectation values that are exponentially suppressed in \(N\) (see Section~\ref{subsec:errormitigation}). 
It has recently been shown that techniques based on fermionic swap networks can
avoid the overheads and disadvantages imposed by the non-locality of the
Jordan-Wigner encoding in a variety of contexts, including during
measurement~\cite{kivlichan2018quantum, motta2018low,o2019generalized}. 
Our work will likewise avoid this challenge without leaving the Jordan-Wigner
framework, allowing estimation of 1- and 2-particle fermionic operator
expectation values by the measurement of only 1-local and 2-local qubit
operators, respectively.

In addition to this reduction in the support of the operators that we measure,
our work offers another opportunity for mitigating errors. 
It has been observed that when one is interested in states with a definite
eigenvalue of a symmetry operator, such as the total particle number, \(\eta\),
or the $z$-component of spin, \(S_z\), it can be useful to have a method which
removes the components of some experimentally prepared state with support on the
wrong symmetry manifold~\cite{mcardle2019error, bonet2018low, o2019calculating,
  mcclean2019decoding}. 
Two basic strategies to accomplish this have been proposed. 
The first of these strategies is to directly and ``non-destructively'' measure
the symmetry operator and discard those outcomes where the undesired eigenvalue
is observed, projecting into the proper symmetry sector by postselection. 
In order to construct efficient measurement schemes, prior work in this
direction has focused on measuring the parities of \(\eta\) and \(S_z\), rather
than the full symmetry operators,~\cite{mcardle2019error, bonet2018low}.
These proposals involve non-local operations that usually require \(O(N)\)
depth, which may induce further errors during their implementation. 
The second class of strategies builds upon the foundation of
Ref.~\citenum{mcclean2017hybrid} and uses additional measurements together with
classical post-processing to calculate expectation values of the projected state
without requiring additional circuit depth~\cite{bonet2018low, o2019calculating,
  mcclean2019decoding}, a procedure which can be efficiently applied to the
parity of the number operator in each spin sector. 
In this work, we show how our proposal for measurement naturally leads to the
ability to postselect directly on the proper eigenvalues of the operators
\(\eta\) and \(S_z\), rather than on their parities.

\section{Results}
\subsection{Using Hamiltonian Factorization for Measurements}

The crux of our strategy for improving the efficiency and error resilience of
Hamiltonian averaging is the application of tensor factorization techniques to
the measurement problem. Using a representation discussed in the context of
quantum computing in
Refs.~\citenum{poulin2014trotter,berry2019qubitization,motta2018low}, we begin
with the factorized form of the electronic structure Hamiltonian in second
quantization:
\begin{equation}
  H = U_0 \left(\sum_{p} g_{p} n_{p} \right) U_0^\dagger + \sum_{\ell=1}^L
  U_\ell \left(\sum_{p q} g_{p q}^{(\ell)} n_{p} n_{q}\right) U_\ell^\dagger,
  \label{eq:decomposed_ham}
\end{equation} where the values $g_p$ and $g_{pq}^{(\ell)}$ are scalars, $n_p =
a^\dagger_p a_p$, and the \(U_\ell\) are unitary operators which implement a
single particle change of orbital basis. Specifically,
\begin{align}
    U = \textrm{exp}\left(\sum_{pq} \kappa_{pq} a^\dagger_p a_q\right), \quad U
    a_p^\dagger U^\dagger & = \sum_q \left[e^{\kappa}\right]_{pq} a_q^\dagger,
\end{align} where $[e^{\kappa}]_{pq}$ is the $p,q$ entry of the matrix
exponential of the anti-Hermitian matrix $\kappa$ that characterizes $U$.

Numerous approaches that accomplish this goal exist, including the density
fitting approximation~\cite{whitten1973coulombic, aquilante2010molcas}, and a
double factorization which begins with a Cholesky decomposition or
eigendecomposition of the two-electron integral tensor~\cite{Pedersen2009-xc,
  motta2018low, beebe1977simplifications, koch2003reduced, aquilante2010molcas,
  purwanto2011assessing, mardirossian2018lowering,Peng2017-oo}. 
In this work, we use such an eigendecomposition and refer readers to \app{low_rank_decomp} and to Refs.~\citenum{motta2018low} and
\citenum{berry2019qubitization} for further details. 
The eigendecomposition step permits discarding small eigenvalues to yield a
controllable approximation to the original Hamiltonian.
While such low rank truncations are not central to our approach and would not
significantly reduce the number of measurements, doing so would asymptotically
reduce $L$ (and thus ultimately, the number of distinct measurement term
groupings). 
Such decompositions have been explored extensively in the context of electronic
structure on classical computers on a far wider range of systems than those
considered here~\cite{Roeggen1986-ib, koch2003reduced, Roeggen2008-bt,
  Boman2008-mm, Pedersen2009-xc, Peng2017-oo}. 
It has been found that $L = O(N)$ is sufficient for the case of arbitrary basis
quantum chemistry, both in the large system and large basis set
limits~\cite{Pedersen2009-xc}. 
Furthermore, specific basis sets exist where $L = 1$, such as the plane wave
basis or dual basis of Ref.~\citenum{babbush2018low}.

Our measurement strategy, which we shall refer to as ``Basis Rotation
Grouping,'' is to apply the $U_\ell$ circuit directly to the quantum state prior
to measurement. This allows us to simultaneously sample all of the $\avg{n_p}$
and $\avg{n_p n_q}$ expectation values in the rotated basis. We can then
estimate the energy as
\begin{equation} \avg{H} = \sum_{p} g_{p} \avg{n_{p}}_0 + \sum_{\ell=1}^L
\sum_{p q} g_{p q}^{(\ell)} \avg{n_{p} n_{q}}_\ell ,
  \label{eq:decomposed_ham_ev}
\end{equation}
where the subscript $\ell$ on the expectation values denotes that they are
sampled after applying the basis transformation $U_\ell$. 
The reason that the $\avg{n_p}_\ell$ and $\avg{n_p n_q}_\ell$ expectation values
can be sampled simultaneously is because under the Jordan-Wigner transformation,
$n_p = (\openone + Z_p)/2$, which is a diagonal qubit operator. 
In practice, we assume a standard measurement in the computational basis, giving
us access to measurement outcomes for all diagonal qubit operators
simultaneously. 
Thus, our approach is able to sample all terms in the Hamiltonian with only $L +
1 = O(N)$ distinct term groups.

Fortunately, the $U_\ell$ are exceptionally efficient to implement, even on
hardware with minimal connectivity. Following the strategy described in
Ref.~\citenum{kivlichan2018quantum}, and assuming that the system is an
eigenstate of the total spin operator, any change of single-particle basis can
be performed using $N^2 / 4 - N / 2$ two qubit gates and gate depth of exactly
$N$, even with the connectivity of only a linear array of
qubits~\cite{kivlichan2018quantum}. This gate depth can actually be improved to
$N / 2$ by further parallelizing the approach of \cite{kivlichan2018quantum},
making using ideas that are explained in the context of multiport interferometry
in~\cite{clements2016optimal}. In fact, a further optimization is possible by
performing the second matrix factorization discussed in
Ref.~\citenum{motta2018low}. This would result in only $O(\log^2 N)$ distict
values of the $g_{pq}^{(\ell)}$ and a gate complexity for implementing the
$U_\ell$ which is reduced to $O(N \log N)$; however, we note that this scaling
is only realized in fairly large systems when $N$ is growing towards the
thermodynamic (large system) rather than continuum (large basis) limit.

The primary objective of our measurement strategy is to reduce the time required
to measure the energy to within a fixed accuracy. 
Because different hardware platforms have different repetition rates, we focus
on quantifying the time required in terms of the number of circuit repetitions. 
We shall present data for electronic ground states that demonstrates the
effectiveness of our Basis Rotation Grouping approach in comparison to three
other measurement strategies and the upper bound of \eq{L1_bound}. 
All calculations were performed using the open source software packages
OpenFermion and Psi4~\cite{mcclean2017openfermion, parrish2017psi4}.
Specifically, we used exact calculations of the variance of expectation values
with respect to the full configuration interaction (FCI) ground state to
determine the number of circuit repetitions required. 
The calculations presented here are performed for symmetrically stretched
Hydrogen chains with various bond lengths and numbers of atoms, for a
symmetrically stretched water molecule, and for a stretched Nitrogen dimer, all
in multiple basis sets.
We justify our focus on the electronic ground states here by noting that most
variational algorithms for chemistry attempt to optimize ansatze that are
already initialized near the ground state. 
For reference, we provide analogous data calculated with respect to the
Hartree-Fock state in the supplementary materials (see \app{data_desc}). 

\begin{figure*}[t!]
  \includegraphics[width= 1.0 \textwidth]{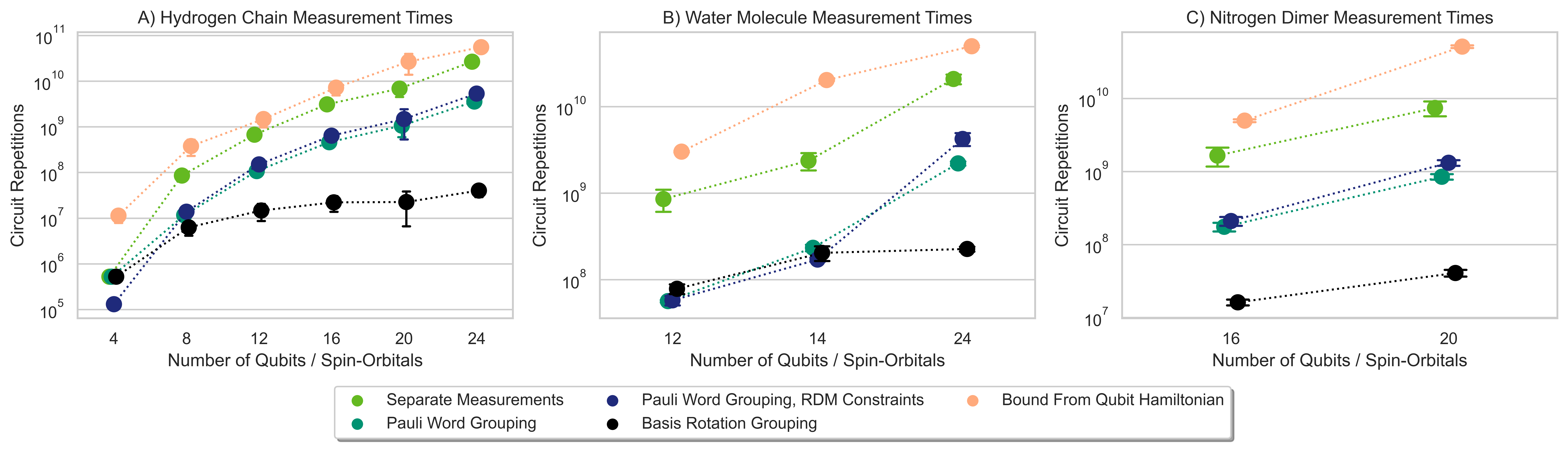}
  \caption{The number of circuit repetitions required to estimate the ground
    state energy of various Hydrogen chains, a water molecule, and a Nitrogen
    dimer with each of the five measurement strategies indicated in the legend. 
    The specific systems considered are enumerated in \tab{system_list}. 
    A target precision corresponding to a \(2\sigma\) error bar of \(1.0\)
    millihartree is assumed. 
    Calculations performed on systems which require the same number of qubits
    (spin-orbitals) are plotted together in columns. 
    The cost of our proposed measurement strategy appears to have a lower
    asymptotic scaling than any other method we consider and obtains a speedup
    of more than an order of magnitude compared to the next best approach for a
    number of systems.}
  \label{fig:measurement_plots}
\end{figure*}

In order to calculate the variance of the estimator of the expectation value of
the energy, it is necessary to determine the distribution of measurements
between the different term groupings. Refs.~\citenum{wecker2015progress}
and~\citenum{rubin2018application} provide a prescription for the optimal
choice. They demand that (in the notation of \eq{L1_bound}) each term $H_\ell$
is measured a fraction of the time $f_\ell$ equal to
\begin{equation}
\label{eq:fraction}
  f_\ell = \frac{\left|\omega_\ell\right| \sqrt{1 - \avg{H_\ell}^2}}{\sum_{j}
  \left|\omega_j\right| \sqrt{1 - \avg{H_j}^2}}.
\end{equation} In practice, the expectation values in the above expression are
not known ahead of time and so the optimal measurement fractions \(f_\ell\)
cannot be efficiently and exactly determined a priori. For the purposes of this
paper, we approximated the ideal distribution of measurements by first
performing a classically tractable configuration interaction singles and doubles
(CISD) calculation of the quantities in \eq{fraction}. We shall show that this
approximation introduces a negligible overhead in measurement time for all
systems considered in this work. One could also 
%imagine 
envisage using an adaptive
measurement scheme that makes additional measurements based on the observed sample
variance, in order to approximate the ideal partitioning of measurement time, such
as the one described in Ref~\citenum{Huggins2019-gi}.

\subsection{Circuit Repetitions Required for Energy Measurement}
\label{subsec:circ_reps}

\begin{table}[t]
\begin{tabular}{l | c | c | c | c}
System & Interatomic & Basis & Frozen & Number of \\
 & Spacings (\AA) & Set & Orbitals & Qubits
\\\hline\hline
H\(_2\) & .6, .7, \ldots 1.3 & STO-3G & None & 4 \\
H\(_2\) & .6, .7, \ldots 1.3 & 6-31G & None & 8 \\
H\(_4\) & .6, .7, \ldots 1.3 & STO-3G & None & 8 \\
H\(_6\) & .6, .7, \ldots 1.3 & STO-3G & None & 12 \\
H\(_4\) & .6, .7, \ldots 1.3 & 6-31G & None & 16 \\
H\(_8\) & .6, .7, \ldots 1.3 & STO-3G & None & 16 \\
H\(_2\) & .6, .7, \ldots 1.3 & cc-pVDZ & None & 20 \\
H\(_{10}\) & .6, .7, \ldots 1.3 & STO-3G & None & 20 \\
H\(_6\) & .6, .7, \ldots 1.3 & 6-31G & None & 24 \\
H\(_2\)O & .8, .9, \ldots 1.5 & STO-3G & 1 & 12 \\
H\(_2\)O & .8, .9, \ldots 1.5 & STO-3G & None & 14 \\
H\(_2\)O & .8, .9, \ldots 1.5 & 6-31G & 1 & 24 \\
N\(_2\) & .9, 1.0, \ldots 1.6 & STO-3G & 2 & 16 \\
N\(_2\) & .9, 1.0, \ldots 1.6 & STO-3G & None & 20 \\

\hline
\end{tabular}
\caption{ List of the molecular systems considered in this work, displayed in
  order of increasing number of qubits, for each type of system. 
  The hydrogen systems consist of a chain of atoms arranged in a line, with
  equal interatomic spacing. 
  The interatomic spacing for the water molecules refers to the length of the
  symmetrically stretched bonds O-H bonds, which are separated by a fixed angle
  of \(104.5\deg\). 
  The active space used for each system has one spatial orbital for every two
  qubits. 
  A non-zero number of frozen orbitals indicates the number of molecular
  orbitals fixed in a totally occupied state.}
\label{tab:system_list}
\end{table}

In \fig{measurement_plots} we plot the number of circuit repetitions for our
proposed ``Basis Rotation Grouping'' measurement approach (black circles),
together with three other measurement strategies and the upper bound based on
\eq{L1_bound} for the systems listed in \tab{system_list}. 
The first and most basic alternative strategy is simply to apply no term
groupings and measure each Pauli string independently, a strategy we refer to as
``Separate Measurements'' (lime green circles). 
A more sophisticated approach, similar to the one described in
Ref.~\citenum{1907.03358}, is to partition the Pauli strings into groups of
terms that can be measured simultaneously. 
In the context of a near-term device we consider two Pauli strings \(P_j\) and
\(P_k\) simultaneously measurable if and only if they act with the same Pauli
operator on all qubits on which they both act non-trivially.
Pauli Strings that satisfy this condition can be simultaneously measured using
only single qubit rotations and measurement.
In order to efficiently partition the Pauli strings into groups we choose to
take all of the terms which only contain \(Z\) operators as one partition and
then account for the remaining Pauli words heuristically by adding them at
random to a group until no more valid choices remain before beginning a new
group. 
We refer to this approach as ``Pauli Word Grouping'' (teal circles). 
The final strategy that we compare with preprocesses the Hamiltonian by applying
the techniques based on the fermionic marginal (RDM) constraints described in
Ref.~\citenum{rubin2018application}, before applying the Jordan-Wigner
transformation and using the same heuristic grouping strategy to group
simultaneously measurable Pauli strings together~\footnote{In the process of
  preparing this manuscript we have become aware of several recent works that
  employ more sophisticated strategies for grouping Pauli words together or
  employing a different family of unitary transformations than those we consider
  to enhance the measurement process~\cite{1907.03358, 1907.07859, 1907.09386,
    1907.09040}. 
  It would be an interesting subject of future work to calculate and compare the
  number of circuit repetitions required by these approaches.}. 
We call this latter strategy ``Pauli Word Grouping, RDM Constraints'' (dark blue
circles).

We refer to the bound of \eq{L1_bound} as being based on the Hamiltonian
coefficients and calculate it from the Jordan-Wigner transformed Hamiltonian,
(meaning that the $\omega_\ell$ in \eq{L1_bound} are the coefficients of Pauli
strings). 
This bound is indicated by salmon-colored circles in \fig{measurement_plots}. 
We note that attempting to calculate a similar bound directly from the fermionic
Hamiltonian (meaning that the $\omega_\ell$ in \eq{L1_bound} would be the
coefficients of the terms $a^\dagger_p a_q$ or $a^\dagger_p a_q^\dagger a_r
a_s$) leads to different bounds. 
These are derived in \app{variance_bounds}, where they are shown to be
substantially looser for the systems we consider in this work. 
While one would not measure the fermion operators directly, it is surprising
that these bounds would be significantly different. 
We refer the interested reader to \app{variance_bounds} for an analysis
and discussion of this phenomenon.

Considering first the Hydrogen chain systems in \fig{measurement_plots} (left
panel, A), we note that our Basis Rotation Grouping approach consistently
outperforms the other strategies for simulations with more than four fermionic
modes, requiring significantly fewer measurements. 
Interestingly, while the bounds from the qubit Hamiltonian and other three
methods appear to have relative performances that are stable across a variety of
system sizes, the Basis Rotation Grouping method appears to have a different
asymptotic scaling, at least for Hydrogen chains of increasing length and basis
set size. 
This is likely due to large scale effects that only manifest when approaching a
system's thermodynamic limit (which one approaches particularly quickly for
Hydrogen chains)~\cite{Simons_collab_2017}. 
In \hyperref[tab:hydrogenAsymptotics]{Table II} we quantify this asymptotic
scaling by assuming that the dependence of the variance on the number of qubits
$N$ in the Hydrogen chain's Hamiltonian can be modeled by the functional form
\(a N^b\) for some constants \(a\) and \(b\) which we fit using a Bayesian
analysis described in the table caption~\cite{granade2012robust}. 
By contrast, the data from the minimal basis water molecule (panel B in
\fig{measurement_plots}) shows no benefit in measurement time from our method
compared to the heuristic grouping strategies.
However, the advantage of our approach becomes significant for that system in
larger basis sets, a trend which is also apparent to a lesser extent for the
Nitrogen dimer (panel C in \fig{measurement_plots}).

\begin{table}[t]
%\begin{tabular}{p {3.5cm} | p {1.1cm} | p {1 cm} | p {1.2 cm} | p {1 cm}}
\begin{tabular}{l | c | c | c | c}
Measurement Strategy & $\avg{\log(a)}$ & $\Delta(a)$ & $\avg{\textbf{b}}$ &
$\sigma(b)$ \\ \hline\hline Bound from Qubit Hamiltonian & -6.0 & 0.3 &
\textbf{4.90} & 0.02 \\ Separate Measurements & -9.3 & 0.4 & \textbf{5.70} &
0.06 \\ Pauli Word Grouping & -8.9 & 0.4 & \textbf{4.88} & 0.06 \\ RDM Constraints & -10.8 & 0.4 & \textbf{5.63} & 0.06\\ Basis Rotation
Grouping & -6.0 & 0.3 & \textbf{2.75} & 0.01 \\ \hline
\end{tabular}
\caption{Bounds and uncertainties result from Bayesian inference using a
  Monte-Carlo approximation with $10^6$ particles for all Hydrogen FCI
  data~\cite{granade2012robust}. 
  We assume $\log(N_{\rm meas}) = \log(a) +\hat{x} + b \log(N)$ where
  $\hat{x}\sim \mathcal{N}(0,0.1)$~\cite{endnote}. 
  The prior distributions are uniform for $\log(a)$ and $b$ over $[-15,1]$ and
  $[1,20]$ respectively. 
  Here $\sigma(b)$ is the posterior standard deviation for $b$ and $\Delta(a)$
  is the posterior standard deviation of $\log(a) + \hat{x}$. 
  "RDM Constraints" refers to the Pauli Word Grouping approach with the RDM
  constraints applied, as in the text.}
\label{tab:hydrogenAsymptotics}
\end{table}

We find that applying the RDM constraints of Ref.~\citenum{rubin2018application}
to our Pauli Word Grouping strategy (the combination is plotted with dark blue
circles in \fig{measurement_plots}) does not significantly reduce the observed
variance, despite the fact that the use of the RDM constraints have been
previously shown to dramatically reduce the bounds on the number of circuit
repetitions required \cite{rubin2018application}. 
In \app{qubit_rdm}, we explore the possibility that this is due to
the fact that these constraints were applied to minimize a bound of the same
form as \eq{L1_bound} that is however formulated using the fermionic
representation of the Hamiltonian. 
We present evidence in \app{variance_bounds} that, in the context of
such bounds, the use of the Jordan-Wigner transformed operators leads to
surprisingly different results. 
However, as we show in \app{qubit_rdm}, we find that the actual variance with respect to the
ground state is not substantially changed by applying the same constraints and
performing the minimization using the qubit representation of the Hamiltonian.

\begin{figure}[t]
  \includegraphics[width= .48\textwidth]{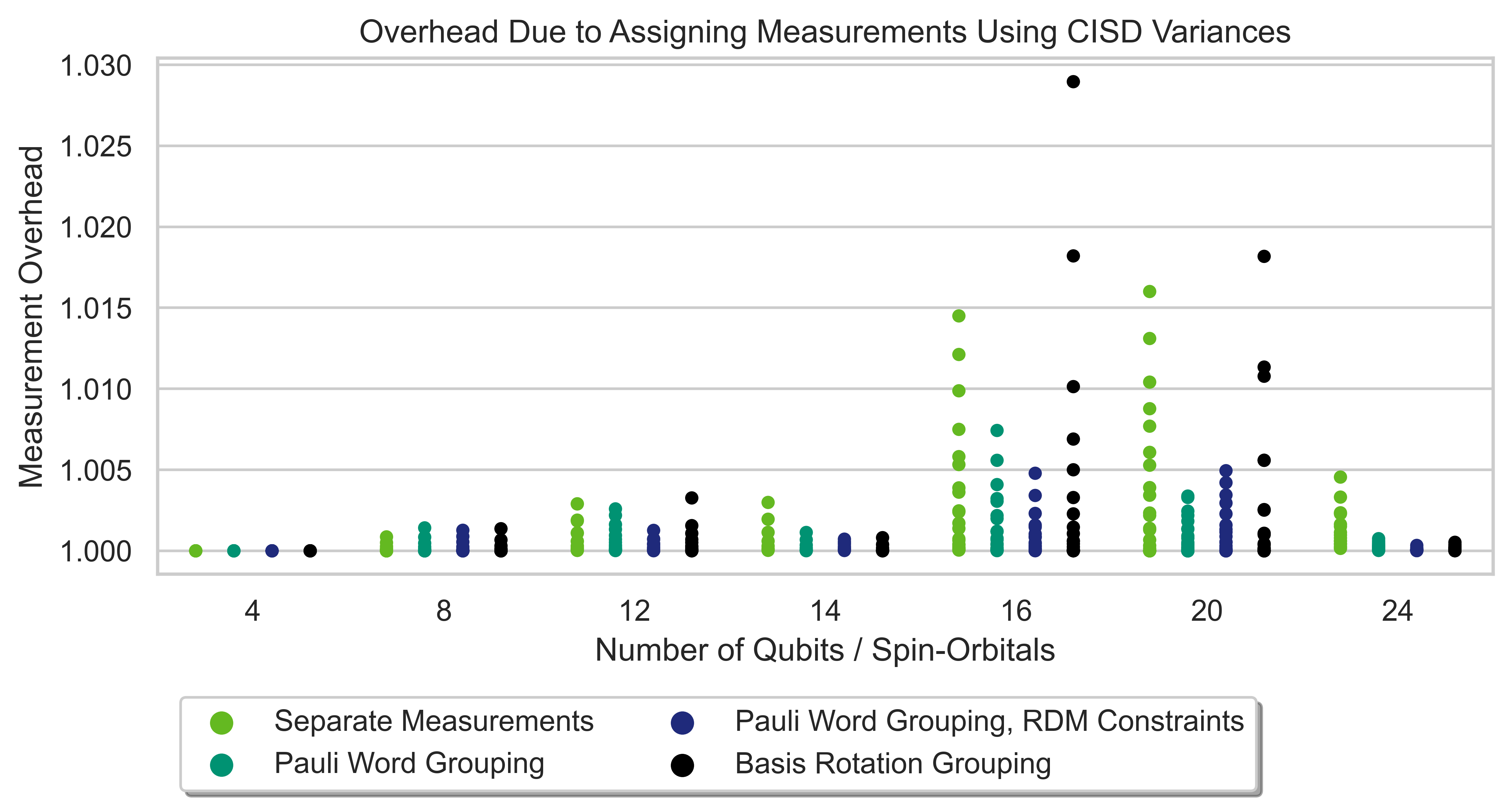}
  \caption{ The increase in the time (or number of circuit repetitions) required
    to measure the ground state energy to a fixed precision when the
    measurements are distributed between groups using the variances calculated
    with the configuration interaction singles and doubles (CISD) approximation
    rather than the true ground state. 
    For each of the systems and measurement techniques considered in this work,
    we present the ratio of the time required when using this approximate
    distribution of measurement repetitions compared with the time required
    using the optimal distribution, both calculated using \eq{fraction} and then
    applied to the measurement of the actual ground state of the system. 
    We find that using a classically tractable CISD calculation to determine the
    distribution of measurements between groups results in only a small increase
    in total measurement time. 
  }
  \label{fig:approximate_variances}
\end{figure}

Earlier we explained that the data presented in \fig{measurement_plots} was
calculated by distributing the measurements between different term groupings
according \eq{fraction} using the variance of each term calculated with a
classically efficient CISD approximation to the ground state. Any deviation from
the ideal allocation of measurement cycles (obtained by evaluating \eq{fraction}
with respect to the true ground state) must increase the time required for
measurement. In \fig{approximate_variances} we present the ratio between the
time required with the approximate distribution and the time required under the
optimal one for each of the systems treated in the work. We find that impact
from this approximation is negligible, with the largest observed increase in
measurement time being below \(3\%\).

Overall, \fig{measurement_plots} speaks for itself in showing that in most cases
there is a very significant reduction in the number of measurements required
when using our strategy - sometimes by up to three orders of magnitude for even
modestly sized systems. Furthermore, these improvements become more significant
as system size grows.

\subsection{Error Mitigation}
\label{subsec:errormitigation}
Beyond the reduction in measurement time, our approach also provides two
distinct forms of error mitigation. 
First, it reduces the susceptibility to readout errors by replacing the
measurement of \(O(N)\) qubit operators with 1 and 2 qubit operators. 
Second, it allows us to perform postselection based on the eigenvalues of the
particle number operators in each spin sector. 
Both properties stem from measuring the Hamiltonian only in terms of density
operators in different basis sets.

The first benefit, the reduction in readout errors, is a consequence of only
needing to measure expectation values of operators that have support on one or
two qubits.
Direct measurement of the Jordan-Wigner transformed Hamiltonian using only
single-qubit rotations and measurement involves measuring operators with support
on \(O(N)\) qubits. 
To demonstrate how reducing the support of the operators helps to mitigate
errors, we consider a simple model of measurement error: the independent,
single-qubit symmetric bitflip channel. 
When estimating the expectation value of a Pauli string \(P_\ell\) acting on $K$
qubits with a single-qubit bitflip error rate $p$, a simple Kraus operator analysis shows that \(P_\ell\) is modified to
\begin{equation}
 \avg{P_\ell}_{\rm bitflip} =
  (1 - 2p)^K \avg{P_\ell}_{\rm true},
\end{equation} which means that the noise channel will bias the estimator of the
expectation value towards zero by a factor exponential in $K$. Thus, the
determination of expectation values is highly sensitive to the extent of locality of the
$P_\ell$, a behavior that we expect to persist under more realistic models of
readout errors.

One could also accomplish the reduction in the support of the operators that our
method achieves by other means. 
For example, one could measure each of the \(O(N^4)\) terms separately,
localizing each one to a single qubit operator by applying \(O(N)\) two-qubit
gates. 
Other schemes have been proposed which simultaneously allow generic two-electron
terms to be measured using \(O(1)\) qubits each while simultaneously
accomplishing the parallel measurement of \(O(N)\) terms at a time, at the cost
of using \(O(N^2)\) or \(O(N^2log(N))\) two-qubit gates~\cite{1907.09040,
  Gokhale2019-cv, o2019generalized}. 
One unique advantage of our approach is that we achieve this reduction in
operator support at the same time as the large reduction in the number of
measurement repetitions presented in Sec.~\ref{subsec:circ_reps} above.

Our approach also enables a second form of error mitigation. 
Each measurement we prescribe is also simultaneously a measurement of the total
particle number operator, \(\eta\), and of the $z$ component of spin, \(S_z\). 
We can therefore reduce the impact of circuit and measurement errors by
performing postselection conditioned on a desired combination of quantum numbers
for each of these operators. 
Let \(P\) denote the projector onto the corresponding subspace and let $\rho$
denote the density matrix of our state. 
We obtain access to the projected expectation value,
\begin{equation}
 \avg{H}_{\rm proj} = \frac{{\rm Tr}\left(P \rho H \right)}{{\rm Tr}\left(P
 \rho\right)},
\end{equation}
directly from the experimental measurement record by discarding those data
points which fall outside of the desired subspace. 
The remaining data points are used to evaluate the expectation values of the
desired Pauli strings.

\begin{figure*}
  \includegraphics[width= 1.0
  \textwidth]{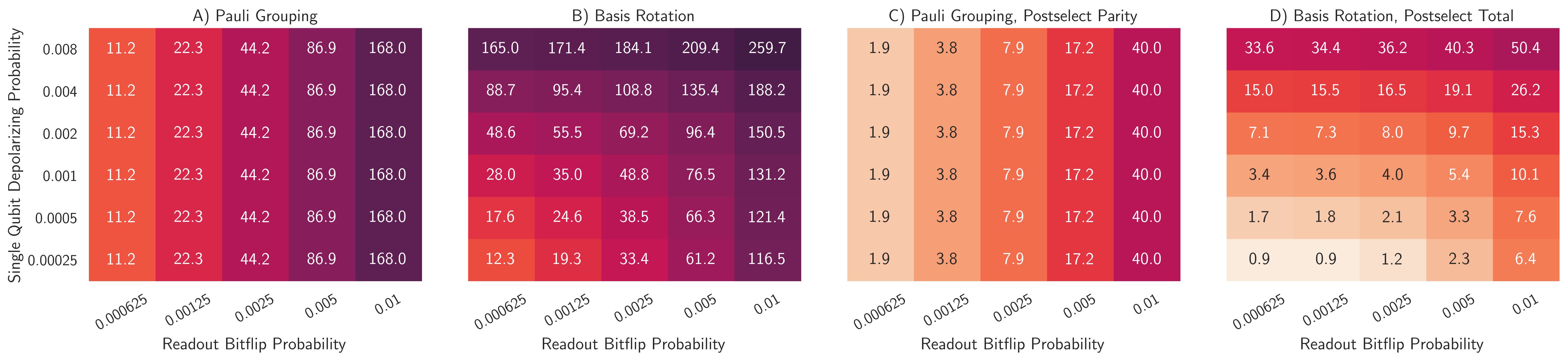}
  \caption{ The absolute error in millihartrees of ground state measurements of
    a stretched chain of six Hydrogen atoms under an error model composed of
    single qubit dephasing noise applied after every two qubit gate together
    with a symmetric bitflip channel during readout. 
    We consider single qubit depolarizing noise with probabilities ranging from
    \(2.5 \times 10^4\) to \(8 \times 10^3\), corresponding to two qubit gate
    error rates of \(\approx 5 \times 10^4\) to \(\approx 1.6 \times 10^2\). 
    For the measurement noise, we take the single qubit bitflip error
    probabilities to be between \(6.25 \times 10^4\) and \(1 \times 10^2\).
    From left to right: A) The error incurred by a ``Pauli Grouping''
    measurement strategy involving simultaneously measuring compatible Pauli
    words in the usual molecular orbital basis. 
    B) The error when using our ``Basis Rotation Grouping'' scheme which
    performs a change of single-particle basis before measurement. 
    C) The errors using the same Pauli word grouping strategy together with
    additional measurements and post-processing which effectively project the
    measured state onto a manifold with the correct parities of the total
    particle number and $S_z$ operators. 
    D) Those found when using our basis rotation strategy and postselecting on
    outcomes where the correct particle number and \(S_z\) were observed. 
    In all panels we consider the measurement of the exact ground state without
    any error during state preparation.
  }
  \label{fig:H6_sim_heatmaps_perfect_state}
\end{figure*}

This postselection is efficient in the sense that it require no additional
machinery beyond what we have already proposed. 
The only cost is a factor of \(\approx 1/ {\rm Tr}(P \rho)\) additional
measurements. 
This factor is approximate because discarding measurements with the wrong
particle number is likely to lead to a lower observed variance. 
Specifically, by removing measurements in the wrong particle number sector, we
avoid having to average over large fluctuations caused by the energetic effects
of adding or removing particles. 
This therefore presents an additional route by which our Basis Rotation
Grouping scheme will reduce the number of measurements in practice. 

Several recent works have proposed error mitigation strategies which allow for
the targeting of specific symmetry sectors. 
We make a brief comparative review of these here in order to place our work in
context. 
One class of strategies focuses on non-destructively measuring one or more
symmetry operators~\cite{mcardle2019error, bonet2018low}.
After performing the measurements and conditioning on the desired eigenvalues,
the post-measurement state becomes $ P \rho P / {\rm Tr}(P \rho)$ and the usual
Hamiltonian averaging can be performed. 
These approaches share some features with our strategy in that they also require
an additional number of measurements that scale as \(1/{\rm Tr}(P \rho)\) and an
increased circuit depth. 
However, they also have some drawbacks that we avoid. 
Because they separate the measurement of the symmetry operator from the
measurement of the Hamiltonian they require the implementation of relatively
complicated non-destructive measurements. 
As a consequence, existing proposals focus on measuring only the parity of the
\(\eta\) and \(S_z\) operators, leading to a strictly less powerful form of
error mitigation than the approach we propose. 
Additionally, most errors that occur during or after the symmetry operator
measurement are undetectable, including errors incurred during readout.

A different class of approaches avoids the need for additional circuit depth at
the expense of requiring more measurements~\cite{bonet2018low, o2019calculating,
mcclean2019decoding}. To understand this, let \(\Pi\) denote the fermionic
parity operator and \(P = (1 + \Pi) / 2\) the projector onto the \(+1\) parity
subspace. Then,
\begin{equation}
  \avg{H}_{\rm proj} = \frac{{\rm Tr}\left(P \rho H\right)}{{\rm Tr}\left(P
  \rho\right)} = \frac{{\rm Tr}\left(\rho H\right) + {\rm Tr}\left(\rho \Pi
  H\right)}{1 + {\rm Tr}\left(\rho \Pi\right)}.
\end{equation} To construct the projected energy it then suffices to measure the
expectation values of the Hamiltonian, the parity operator, and the product of
the Hamiltonian and parity operators. A stochastic sampling scheme and a careful
analysis of the cost of such an approach reveals that it is possible to use
post-processing to estimate the projection onto the subspace with the correct
particle number parity in each spin sector at a cost of roughly \(1/ {\rm
Tr}(P_{\uparrow} P_{\downarrow} \rho)^2\) (where \(P_\uparrow\) and
\(P_\downarrow\) are the parity projectors for the two spin
sectors)~\cite{mcclean2019decoding}. Unlike our approach, this class of error
mitigation techniques does not easily allow for the projection onto the correct
eigenvalues of $\eta$ and \(S_z\), owing to the large number of terms required
to construct these projection operators. Furthermore, the scaling in the number
of additional measurements we described above, already more costly than our
approach, is also too generous. This is because the product of the parity
operators and the Hamiltonian will contain a larger number non-simultaneously
measurable terms than the same Hamiltonian on its own. Maximum efficiency may
require grouping schemes that consider this larger number of term groupings.

\begin{figure*}
  \includegraphics[width= 1.0 \textwidth]{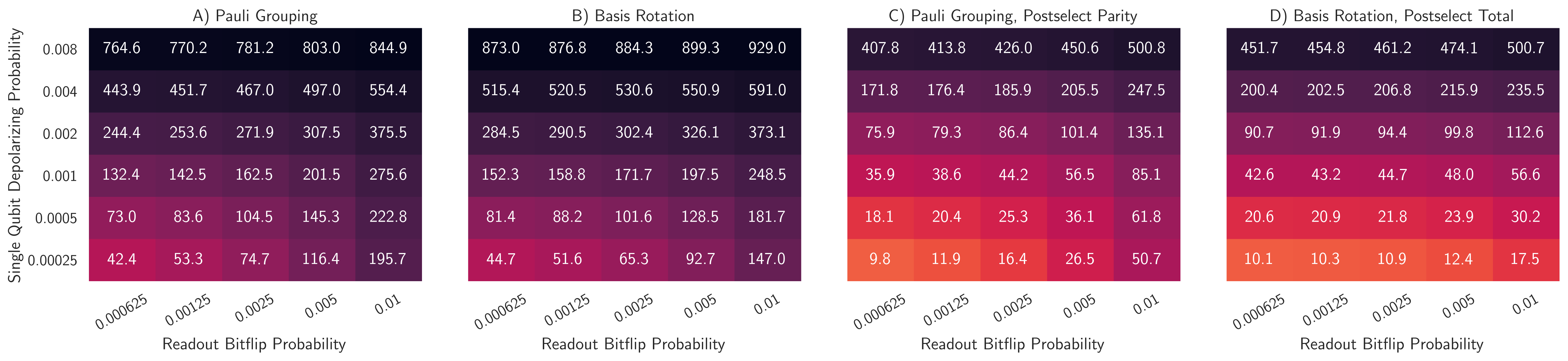}
  \caption{ The absolute error in millihartrees of ground state measurements of
    a stretched chain of six Hydrogen atoms under an error model composed of
    single qubit dephasing noise applied after every two qubit gate together
    with a symmetric bitflip channel during readout. 
    We consider single qubit depolarizing noise with probabilities ranging from
    \(2.5 \times 10^4\) to \(8 \times 10^3\), corresponding to two qubit gate
    error rates of \(\approx 5 \times 10^4\) to \(\approx 1.6 \times 10^2\). 
    For the measurement noise, we take the single qubit bitflip error
    probabilities to be between \(6.25 \times 10^4\) and \(1 \times 10^2\).
    From left to right: A) The error incurred by a ``Pauli Grouping''
    measurement strategy involving simultaneously measuring compatible Pauli
    words in the usual molecular orbital basis. 
    B) The error when using our ``Basis Rotation Grouping'' scheme which
    performs a change of single-particle basis before measurement. 
    C) The errors using the same Pauli word grouping strategy together with
    additional measurements and post-processing which effectively project the
    measured state onto a manifold with the correct parities of the total
    particle number and $S_z$ operators. 
    D) Those found when using our basis rotation strategy and postselecting on
    outcomes where the correct particle number and \(S_z\) were observed. 
    In all panels, for the purpose of approximating a realistic ansatz circuit,
    three random Givens rotation networks which compose to the identity were
    simulated acting on the ground state prior to measurement.
  }
  \label{fig:H6_sim_heatmaps_state_prep}
\end{figure*}

The most significant drawback of our method in the context of error mitigation
is that the additional time and gates required for the basis transformation
circuit lead to additional opportunities for errors. 
We believe the reduction in circuit repetitions we have shown makes our method
the most attractive choice when it is feasible to use an additional \(O(N^2)\)
two-qubit gates during the measurement process. 
We therefore focus on comparing the performance of our strategy with a strategy
that requires no additional gates and uses a quantum subspace error mitigation
approach that effectively projects onto the correct parity of the number
operator on each spin sector~\cite{bonet2018low, mcclean2019decoding}. 
In order to do so, we use the open source software package Cirq~\cite{cirq_2019}
to simulate the performance of both strategies for measuring the ground state
energy of a chain of six Hydrogen atoms symmetrically stretched to 1.3\AA~in an
STO-3G basis. 
We take an error model consisting of i) applying a single qubit depolarizing
channel with some probability to both qubits following each two qubit gate, and
ii) applying a bitflip channel during the measurement process with some other
probability. 
We report results for a wide range of gate and readout noise levels inspired by
the capabilities of state of the art superconducting and ion trap quantum
computers~\cite{Kjaergaard2020-bb, Bruzewicz2019-zm, Heinsoo2018-tp,
  Preskill2018-po}. 
Specfically, we consider single qubit depolarizing noise with probabilities
ranging from \(2.5 \times 10^4\) to \(8 \times 10^3\) and single qubit bitflip
error probabilities between \(6.25 \times 10^4\) and \(1 \times 10^2\). 
Here we do not consider the effect of a finite number of measurements and
instead report the expectation values from the final density matrix.

\fig{H6_sim_heatmaps_perfect_state} shows the error in the measurement of the
ground state energy for the error-mitigated Basis Rotation Grouping (far right
panel) and Pauli Word Grouping (second panel from right) approaches together
with the expectation values for both measurement strategies without error
mitigation (two left panels). 
In these calculations we assumed that the ground state wavefunction under the
Jordan-Wigner transformation is prepared without error.
Circuit level noise is considered only during the execution of the Givens
rotation required for our Basis Rotation Grouping approach. 
In order to include the impact of our proposed error mitigation strategy on
state preparation as well as measurement, we have also carried out calculations
including circuit noise during state preparation. 
The results of these calculations are presented in
\fig{H6_sim_heatmaps_state_prep}.
Here we have approximated a realistic state preparation circuit by applying
three random basis rotations which compose to the identity to the ground state
wavefunction. 
These state preparation circuits are simulated with the same gate noise as the
measurement circuits. 
This choice is motivated by the assumption that low-depth circuits will be
required for the successful application of VQE and the expectation that \(90\)
two-qubit gates represents a reasonable lower bound to the size of circuit for a
strongly-correlated problem on \(12\) qubits.

Figures \ref{fig:H6_sim_heatmaps_perfect_state} and
\ref{fig:H6_sim_heatmaps_state_prep} show that the Pauli Word Grouping and Basis
Rotation Grouping approaches to measurement benefit significantly from their
respective error mitigation strategies. 
Despite the fact that our proposed Basis Rotation Grouping technique requires
\(30\) additional two-qubit gates compared to the Pauli Word Grouping approach,
we see that the errors remaining after mitigation are comparable in some regimes
and are lower for our strategy when noise during measurement is the dominant
error channel (compare the bottom right corners of the two rightmost panels in
both figures). 
Focusing first on \fig{H6_sim_heatmaps_perfect_state}, we can see that this is
true even when the errors during state preparation are not taken into account. 
Examining the left two panels of both figures, we can see that even without
applying postselection,the locality of our Jordan-Wigner transformed operators
leads to a considerable benefit in suppressing the impact of readout errors.

We note that the absolute errors we find when including noise during state
preparation (\fig{H6_sim_heatmaps_state_prep}), even at the lowest noise levels
considered here, are larger than the usual target of ``chemical accuracy''
($\sim$ 1 mHa). 
In practice, an experimental implementation of VQE on non-trivial systems will
require the combination of multiple forms of error mitigation. 
Prior work has shown that error mitigation by symmetry projection combines
favorably with proposals to extrapolate expectation values to the zero noise
limit~\cite{mcardle2019error}. 
We expect that such an extrapolation procedure could significantly improve the
numbers we present here. 
Other avenues for potential improvements are also available. 
For example, one could rely on the error mitigation and efficiency provided by
our measurement strategy during the outer loop optimization procedure, before
utilizing a richer quantum subspace expansion in an attempt to reduce errors in
the ground state energy after determining the optimal ansatz parameters.

\section{Discussion}

We have presented an improved strategy for measuring the expectation value of
the quantum chemical Hamiltonian on near-term quantum computers. 
Our approach makes use of well studied factorizations of the two-electron
integral tensor, in order to rewrite the Hamiltonian in a form which is
especially convenient for measuring under the Jordan-Wigner transformation. 
By doing so, we obtain \(O(N)\) distinct sets of terms which must be measured
separately, instead of the \(O(N^4)\) required by a naive counting of terms
approach. 
Application to specific molecular systems show that in practice, we require a
much smaller number of repetitions to measure the ground state energy to within
a fixed accuracy target. 
For example, assuming an experimental repetition rate of 10 kHz (consistent with
the capabilities of commercial superconducting qubit platforms), a commonly
referenced bound based on the Hamiltonian coefficients suggests that
approximately \(55\) days are required to estimate the ground state energy to of
a symmetrically stretched chain of 6 Hydrogen atoms encoded as a wavefunction on
24 qubits to within chemical accuracy, while our approach requires only \(44\)
minutes. 
Our proposed measurement approach also removes the susceptibility to readout
error caused by long Jordan-Wigner strings and allows for postselection by
simultaneously measuring the total particle number and \(S_z\) operators with
each measurement shot.

The tensor factorization that we used to realize our measurement
strategy is only one of a family of such factorizations. Future work might
explore the use of different factorizations, or even tailor the choice of single
particle bases for measurement to a particular system, by choosing them with some
knowledge of the variances and covariances between terms in the Hamiltonian. As
a more concrete direction for future work, the data we show in
\app{variance_bounds} regarding the difference between the bounds when calculated
directly from the fermionic operators and the same approach applied to the
Jordan-Wigner transformed operators, suggests that the cost estimates for
error-corrected quantum algorithms should be recalculated using the qubit
Hamiltonian.

For the largest systems we consider in this work, the 24 qubit Hydrogen chain
and water simulations, and the 20 qubit Nitrogen calculations, our numerical
results indicate that using our approach results in a speedup of more than an
order of magnitude when compared to recent state-of-the-art measurement
strategies. 
Furthermore, we observe a speedup of more than three orders of magnitude
compared to the bounds commonly used to perform estimates in the literature. 
We also present strong evidence for an asymptotic improvement in our data on
Hydrogen chains of various sizes. 
We performed detailed circuit simulations that show that reduction in readout
errors combined with the error mitigation enabled by our work largely balances
out the requirement for deeper circuits, even when compared against a moderately
expensive error mitigation strategy based on the quantum subspace
expansion~\cite{bonet2018low}. 
We expect that the balance of reduced measurement time and efficient error
mitigation provided by our approach will be useful in the application of
variational quantum algorithms to more complex molecular systems.

Finally, we note that these techniques will generally be useful for quantum
simulating any fermionic system, even those for which the tensor factorization
cannot be truncated, such as the Sachdev-Ye-Kitaev model of many-body chaotic
dynamics \cite{kitaev2015syk,babbush2019syk}. 
In that case, \(L\) will attain its maximal value of \(N^2\), and our scheme
will require \(N^2 + 1\) partitions. 
Likewise, if the goal is to use the basis rotation grouping technique to
estimate the fermionic 2-particle reduced density matrix rather than just the
energy, one would need to measure in all $O(N^2)$ bases.

\subsection*{Acknowledgements}

The authors thank Dominic Berry for his insight that the techniques of
\cite{clements2016optimal} can be used to halve the gate depth of the basis
rotation circuits presented in \cite{kivlichan2018quantum}. KBW was supported by
the U.S. Department of Energy, Office of Science, Office of Advanced Scientific
Computing Research, Quantum algorithm Teams Program, under contract number
DE-AC02-05CH11231.

\subsection*{Ethics Declarations} The authors declare no competing interests.

%\subsection*{Author Contribution} W.H. and R.B. conceived the idea and co-wrote
%the majority of the paper. W.H. performed all numerical simulations except for
%the Bayesian analysis, which N.W. carried out. W.H, J.M, N.R, Z.J, N.W, K.B.W.
%and R.B. all participated in discussions which developed the theory and shaped
%the project.

\subsection*{Data Availability} The data that support the findings of this study
are available from the corresponding author upon reasonable request.

\subsection*{Code Availability} Much of the code that support the findings of
this study is already available in the OpenFermion
library~\cite{mcclean2017openfermion}. The remainder is available from the
corresponding author upon reasonable request.

% \bibliography{refs}

\begin{thebibliography}{57}%
\makeatletter
\providecommand \@ifxundefined [1]{%
 \@ifx{#1\undefined}
}%
\providecommand \@ifnum [1]{%
 \ifnum #1\expandafter \@firstoftwo
 \else \expandafter \@secondoftwo
 \fi
}%
\providecommand \@ifx [1]{%
 \ifx #1\expandafter \@firstoftwo
 \else \expandafter \@secondoftwo
 \fi
}%
\providecommand \natexlab [1]{#1}%
\providecommand \enquote  [1]{``#1''}%
\providecommand \bibnamefont  [1]{#1}%
\providecommand \bibfnamefont [1]{#1}%
\providecommand \citenamefont [1]{#1}%
\providecommand \href@noop [0]{\@secondoftwo}%
\providecommand \href [0]{\begingroup \@sanitize@url \@href}%
\providecommand \@href[1]{\@@startlink{#1}\@@href}%
\providecommand \@@href[1]{\endgroup#1\@@endlink}%
\providecommand \@sanitize@url [0]{\catcode `\\12\catcode `\$12\catcode
  `\&12\catcode `\#12\catcode `\^12\catcode `\_12\catcode `\%12\relax}%
\providecommand \@@startlink[1]{}%
\providecommand \@@endlink[0]{}%
\providecommand \url  [0]{\begingroup\@sanitize@url \@url }%
\providecommand \@url [1]{\endgroup\@href {#1}{\urlprefix }}%
\providecommand \urlprefix  [0]{URL }%
\providecommand \Eprint [0]{\href }%
\providecommand \doibase [0]{http://dx.doi.org/}%
\providecommand \selectlanguage [0]{\@gobble}%
\providecommand \bibinfo  [0]{\@secondoftwo}%
\providecommand \bibfield  [0]{\@secondoftwo}%
\providecommand \translation [1]{[#1]}%
\providecommand \BibitemOpen [0]{}%
\providecommand \bibitemStop [0]{}%
\providecommand \bibitemNoStop [0]{.\EOS\space}%
\providecommand \EOS [0]{\spacefactor3000\relax}%
\providecommand \BibitemShut  [1]{\csname bibitem#1\endcsname}%
\let\auto@bib@innerbib\@empty
%</preamble>
\bibitem [{\citenamefont {Peruzzo}\ \emph {et~al.}(2014)\citenamefont
  {Peruzzo}, \citenamefont {McClean}, \citenamefont {Shadbolt}, \citenamefont
  {Yung}, \citenamefont {Zhou}, \citenamefont {Love}, \citenamefont
  {Aspuru-Guzik},\ and\ \citenamefont {O’brien}}]{peruzzo2014variational}%
  \BibitemOpen
  \bibfield  {author} {\bibinfo {author} {\bibfnamefont {Alberto}\ \bibnamefont
  {Peruzzo}}, \bibinfo {author} {\bibfnamefont {Jarrod}\ \bibnamefont
  {McClean}}, \bibinfo {author} {\bibfnamefont {Peter}\ \bibnamefont
  {Shadbolt}}, \bibinfo {author} {\bibfnamefont {Man-Hong}\ \bibnamefont
  {Yung}}, \bibinfo {author} {\bibfnamefont {Xiao-Qi}\ \bibnamefont {Zhou}},
  \bibinfo {author} {\bibfnamefont {Peter~J}\ \bibnamefont {Love}}, \bibinfo
  {author} {\bibfnamefont {Al{\'a}n}\ \bibnamefont {Aspuru-Guzik}}, \ and\
  \bibinfo {author} {\bibfnamefont {Jeremy~L}\ \bibnamefont {O’brien}},\
  }\bibfield  {title} {\enquote {\bibinfo {title} {A variational eigenvalue
  solver on a photonic quantum processor},}\ }\href
  {https://www.nature.com/articles/ncomms5213} {\bibfield  {journal} {\bibinfo
  {journal} {Nat. Commun.}\ }\textbf {\bibinfo {volume} {5}},\ \bibinfo {pages}
  {4213} (\bibinfo {year} {2014})}\BibitemShut {NoStop}%
\bibitem [{\citenamefont {McClean}\ \emph {et~al.}(2016)\citenamefont
  {McClean}, \citenamefont {Romero}, \citenamefont {Babbush},\ and\
  \citenamefont {Aspuru-Guzik}}]{mcclean2016theory}%
  \BibitemOpen
  \bibfield  {author} {\bibinfo {author} {\bibfnamefont {Jarrod~R}\
  \bibnamefont {McClean}}, \bibinfo {author} {\bibfnamefont {Jonathan}\
  \bibnamefont {Romero}}, \bibinfo {author} {\bibfnamefont {Ryan}\ \bibnamefont
  {Babbush}}, \ and\ \bibinfo {author} {\bibfnamefont {Al{\'a}n}\ \bibnamefont
  {Aspuru-Guzik}},\ }\bibfield  {title} {\enquote {\bibinfo {title} {The theory
  of variational hybrid quantum-classical algorithms},}\ }\href
  {https://iopscience.iop.org/article/10.1088/1367-2630/18/2/023023/meta}
  {\bibfield  {journal} {\bibinfo  {journal} {New J. Phys.}\ }\textbf {\bibinfo
  {volume} {18}},\ \bibinfo {pages} {023023} (\bibinfo {year}
  {2016})}\BibitemShut {NoStop}%
\bibitem [{\citenamefont {Wecker}\ \emph {et~al.}(2015)\citenamefont {Wecker},
  \citenamefont {Hastings},\ and\ \citenamefont {Troyer}}]{wecker2015progress}%
  \BibitemOpen
  \bibfield  {author} {\bibinfo {author} {\bibfnamefont {Dave}\ \bibnamefont
  {Wecker}}, \bibinfo {author} {\bibfnamefont {Matthew~B}\ \bibnamefont
  {Hastings}}, \ and\ \bibinfo {author} {\bibfnamefont {Matthias}\ \bibnamefont
  {Troyer}},\ }\bibfield  {title} {\enquote {\bibinfo {title} {Progress towards
  practical quantum variational algorithms},}\ }\href
  {https://journals.aps.org/pra/abstract/10.1103/PhysRevA.92.042303} {\bibfield
   {journal} {\bibinfo  {journal} {Phys. Rev. A}\ }\textbf {\bibinfo {volume}
  {92}},\ \bibinfo {pages} {042303} (\bibinfo {year} {2015})}\BibitemShut
  {NoStop}%
\bibitem [{\citenamefont {O'Malley}\ \emph {et~al.}(2016)\citenamefont
  {O'Malley}, \citenamefont {Babbush}, \citenamefont {Kivlichan}, \citenamefont
  {Romero}, \citenamefont {McClean}, \citenamefont {Barends}, \citenamefont
  {Kelly}, \citenamefont {Roushan}, \citenamefont {Tranter}, \citenamefont
  {Ding}, \citenamefont {Campbell}, \citenamefont {Chen}, \citenamefont {Chen},
  \citenamefont {Chiaro}, \citenamefont {Dunsworth}, \citenamefont {Fowler},
  \citenamefont {Jeffrey}, \citenamefont {Megrant}, \citenamefont {Mutus},
  \citenamefont {Neill}, \citenamefont {Quintana}, \citenamefont {Sank},
  \citenamefont {Vainsencher}, \citenamefont {Wenner}, \citenamefont {White},
  \citenamefont {Coveney}, \citenamefont {Love}, \citenamefont {Neven},
  \citenamefont {Aspuru-Guzik},\ and\ \citenamefont {Martinis}}]{OMalley2016}%
  \BibitemOpen
  \bibfield  {author} {\bibinfo {author} {\bibfnamefont {P~J~J}\ \bibnamefont
  {O'Malley}}, \bibinfo {author} {\bibfnamefont {R}~\bibnamefont {Babbush}},
  \bibinfo {author} {\bibfnamefont {I~D}\ \bibnamefont {Kivlichan}}, \bibinfo
  {author} {\bibfnamefont {J}~\bibnamefont {Romero}}, \bibinfo {author}
  {\bibfnamefont {J~R}\ \bibnamefont {McClean}}, \bibinfo {author}
  {\bibfnamefont {R}~\bibnamefont {Barends}}, \bibinfo {author} {\bibfnamefont
  {J}~\bibnamefont {Kelly}}, \bibinfo {author} {\bibfnamefont {P}~\bibnamefont
  {Roushan}}, \bibinfo {author} {\bibfnamefont {A}~\bibnamefont {Tranter}},
  \bibinfo {author} {\bibfnamefont {N}~\bibnamefont {Ding}}, \bibinfo {author}
  {\bibfnamefont {B}~\bibnamefont {Campbell}}, \bibinfo {author} {\bibfnamefont
  {Y}~\bibnamefont {Chen}}, \bibinfo {author} {\bibfnamefont {Z}~\bibnamefont
  {Chen}}, \bibinfo {author} {\bibfnamefont {B}~\bibnamefont {Chiaro}},
  \bibinfo {author} {\bibfnamefont {A}~\bibnamefont {Dunsworth}}, \bibinfo
  {author} {\bibfnamefont {A~G}\ \bibnamefont {Fowler}}, \bibinfo {author}
  {\bibfnamefont {E}~\bibnamefont {Jeffrey}}, \bibinfo {author} {\bibfnamefont
  {A}~\bibnamefont {Megrant}}, \bibinfo {author} {\bibfnamefont {J~Y}\
  \bibnamefont {Mutus}}, \bibinfo {author} {\bibfnamefont {C}~\bibnamefont
  {Neill}}, \bibinfo {author} {\bibfnamefont {C}~\bibnamefont {Quintana}},
  \bibinfo {author} {\bibfnamefont {D}~\bibnamefont {Sank}}, \bibinfo {author}
  {\bibfnamefont {A}~\bibnamefont {Vainsencher}}, \bibinfo {author}
  {\bibfnamefont {J}~\bibnamefont {Wenner}}, \bibinfo {author} {\bibfnamefont
  {T~C}\ \bibnamefont {White}}, \bibinfo {author} {\bibfnamefont {P~V}\
  \bibnamefont {Coveney}}, \bibinfo {author} {\bibfnamefont {P~J}\ \bibnamefont
  {Love}}, \bibinfo {author} {\bibfnamefont {H}~\bibnamefont {Neven}}, \bibinfo
  {author} {\bibfnamefont {A}~\bibnamefont {Aspuru-Guzik}}, \ and\ \bibinfo
  {author} {\bibfnamefont {J~M}\ \bibnamefont {Martinis}},\ }\bibfield  {title}
  {\enquote {\bibinfo {title} {{Scalable Quantum Simulation of Molecular
  Energies}},}\ }\href {\doibase http://dx.doi.org/10.1103/PhysRevX.6.031007}
  {\bibfield  {journal} {\bibinfo  {journal} {Phys. Rev. X}\ }\textbf {\bibinfo
  {volume} {6}},\ \bibinfo {pages} {31007} (\bibinfo {year}
  {2016})}\BibitemShut {NoStop}%
\bibitem [{\citenamefont {Kandala}\ \emph {et~al.}(2017)\citenamefont
  {Kandala}, \citenamefont {Mezzacapo}, \citenamefont {Temme}, \citenamefont
  {Takita}, \citenamefont {Brink}, \citenamefont {Chow},\ and\ \citenamefont
  {Gambetta}}]{kandala2017hardware}%
  \BibitemOpen
  \bibfield  {author} {\bibinfo {author} {\bibfnamefont {Abhinav}\ \bibnamefont
  {Kandala}}, \bibinfo {author} {\bibfnamefont {Antonio}\ \bibnamefont
  {Mezzacapo}}, \bibinfo {author} {\bibfnamefont {Kristan}\ \bibnamefont
  {Temme}}, \bibinfo {author} {\bibfnamefont {Maika}\ \bibnamefont {Takita}},
  \bibinfo {author} {\bibfnamefont {Markus}\ \bibnamefont {Brink}}, \bibinfo
  {author} {\bibfnamefont {Jerry~M}\ \bibnamefont {Chow}}, \ and\ \bibinfo
  {author} {\bibfnamefont {Jay~M}\ \bibnamefont {Gambetta}},\ }\bibfield
  {title} {\enquote {\bibinfo {title} {Hardware-efficient variational quantum
  eigensolver for small molecules and quantum magnets},}\ }\href
  {https://www.nature.com/articles/nature23879} {\bibfield  {journal} {\bibinfo
   {journal} {Nature}\ }\textbf {\bibinfo {volume} {549}},\ \bibinfo {pages}
  {242} (\bibinfo {year} {2017})}\BibitemShut {NoStop}%
\bibitem [{\citenamefont {Lee}\ \emph {et~al.}(2018)\citenamefont {Lee},
  \citenamefont {Huggins}, \citenamefont {Head-Gordon},\ and\ \citenamefont
  {Whaley}}]{lee2018generalized}%
  \BibitemOpen
  \bibfield  {author} {\bibinfo {author} {\bibfnamefont {Joonho}\ \bibnamefont
  {Lee}}, \bibinfo {author} {\bibfnamefont {William~J}\ \bibnamefont
  {Huggins}}, \bibinfo {author} {\bibfnamefont {Martin}\ \bibnamefont
  {Head-Gordon}}, \ and\ \bibinfo {author} {\bibfnamefont {K~Birgitta}\
  \bibnamefont {Whaley}},\ }\bibfield  {title} {\enquote {\bibinfo {title}
  {Generalized unitary coupled cluster wave functions for quantum
  computation},}\ }\href {https://pubs.acs.org/doi/10.1021/acs.jctc.8b01004}
  {\bibfield  {journal} {\bibinfo  {journal} {J. Chem. Theory Comput.}\
  }\textbf {\bibinfo {volume} {15}},\ \bibinfo {pages} {311--324} (\bibinfo
  {year} {2018})}\BibitemShut {NoStop}%
\bibitem [{\citenamefont {Parrish}\ \emph {et~al.}(2019)\citenamefont
  {Parrish}, \citenamefont {Hohenstein}, \citenamefont {McMahon},\ and\
  \citenamefont {Mart{\'\i}nez}}]{parrish2019quantum}%
  \BibitemOpen
  \bibfield  {author} {\bibinfo {author} {\bibfnamefont {Robert~M}\
  \bibnamefont {Parrish}}, \bibinfo {author} {\bibfnamefont {Edward~G}\
  \bibnamefont {Hohenstein}}, \bibinfo {author} {\bibfnamefont {Peter~L}\
  \bibnamefont {McMahon}}, \ and\ \bibinfo {author} {\bibfnamefont {Todd~J}\
  \bibnamefont {Mart{\'\i}nez}},\ }\bibfield  {title} {\enquote {\bibinfo
  {title} {Quantum computation of electronic transitions using a variational
  quantum eigensolver},}\ }\href
  {https://journals.aps.org/prl/abstract/10.1103/PhysRevLett.122.230401}
  {\bibfield  {journal} {\bibinfo  {journal} {Phys. Rev. Lett.}\ }\textbf
  {\bibinfo {volume} {122}},\ \bibinfo {pages} {230401} (\bibinfo {year}
  {2019})}\BibitemShut {NoStop}%
\bibitem [{\citenamefont {O'Brien}\ \emph {et~al.}(2019)\citenamefont
  {O'Brien}, \citenamefont {Senjean}, \citenamefont {Sagastizabal},
  \citenamefont {Bonet-Monroig}, \citenamefont {Dutkiewicz}, \citenamefont
  {Buda}, \citenamefont {DiCarlo},\ and\ \citenamefont
  {Visscher}}]{o2019calculating}%
  \BibitemOpen
  \bibfield  {author} {\bibinfo {author} {\bibfnamefont {TE}~\bibnamefont
  {O'Brien}}, \bibinfo {author} {\bibfnamefont {B}~\bibnamefont {Senjean}},
  \bibinfo {author} {\bibfnamefont {R}~\bibnamefont {Sagastizabal}}, \bibinfo
  {author} {\bibfnamefont {X}~\bibnamefont {Bonet-Monroig}}, \bibinfo {author}
  {\bibfnamefont {A}~\bibnamefont {Dutkiewicz}}, \bibinfo {author}
  {\bibfnamefont {F}~\bibnamefont {Buda}}, \bibinfo {author} {\bibfnamefont
  {L}~\bibnamefont {DiCarlo}}, \ and\ \bibinfo {author} {\bibfnamefont
  {L}~\bibnamefont {Visscher}},\ }\bibfield  {title} {\enquote {\bibinfo
  {title} {Calculating energy derivatives for quantum chemistry on a quantum
  computer},}\ }\href {http://arxiv.org/abs/1905.03742} {\bibfield  {journal}
  {\bibinfo  {journal} {arXiv:1905.03742}\ } (\bibinfo {year}
  {2019})}\BibitemShut {NoStop}%
\bibitem [{\citenamefont {Bonet-Monroig}\ \emph {et~al.}(2018)\citenamefont
  {Bonet-Monroig}, \citenamefont {Sagastizabal}, \citenamefont {Singh},\ and\
  \citenamefont {O'Brien}}]{bonet2018low}%
  \BibitemOpen
  \bibfield  {author} {\bibinfo {author} {\bibfnamefont {X}~\bibnamefont
  {Bonet-Monroig}}, \bibinfo {author} {\bibfnamefont {R}~\bibnamefont
  {Sagastizabal}}, \bibinfo {author} {\bibfnamefont {M}~\bibnamefont {Singh}},
  \ and\ \bibinfo {author} {\bibfnamefont {TE}~\bibnamefont {O'Brien}},\
  }\bibfield  {title} {\enquote {\bibinfo {title} {Low-cost error mitigation by
  symmetry verification},}\ }\href
  {https://journals.aps.org/pra/abstract/10.1103/PhysRevA.98.062339} {\bibfield
   {journal} {\bibinfo  {journal} {Phys. Rev. A}\ }\textbf {\bibinfo {volume}
  {98}},\ \bibinfo {pages} {062339} (\bibinfo {year} {2018})}\BibitemShut
  {NoStop}%
\bibitem [{\citenamefont {McClean}\ \emph {et~al.}(2020)\citenamefont
  {McClean}, \citenamefont {Jiang}, \citenamefont {Rubin}, \citenamefont
  {Babbush},\ and\ \citenamefont {Neven}}]{mcclean2019decoding}%
  \BibitemOpen
  \bibfield  {author} {\bibinfo {author} {\bibfnamefont {Jarrod~R}\
  \bibnamefont {McClean}}, \bibinfo {author} {\bibfnamefont {Zhang}\
  \bibnamefont {Jiang}}, \bibinfo {author} {\bibfnamefont {Nicholas~C}\
  \bibnamefont {Rubin}}, \bibinfo {author} {\bibfnamefont {Ryan}\ \bibnamefont
  {Babbush}}, \ and\ \bibinfo {author} {\bibfnamefont {Hartmut}\ \bibnamefont
  {Neven}},\ }\bibfield  {title} {\enquote {\bibinfo {title} {Decoding quantum
  errors with subspace expansions},}\ }\href {\doibase
  10.1038/s41467-020-14341-w} {\bibfield  {journal} {\bibinfo  {journal} {Nat.
  Commun.}\ }\textbf {\bibinfo {volume} {11}},\ \bibinfo {pages} {636}
  (\bibinfo {year} {2020})}\BibitemShut {NoStop}%
\bibitem [{\citenamefont {McArdle}\ \emph {et~al.}(2019)\citenamefont
  {McArdle}, \citenamefont {Yuan},\ and\ \citenamefont
  {Benjamin}}]{mcardle2019error}%
  \BibitemOpen
  \bibfield  {author} {\bibinfo {author} {\bibfnamefont {Sam}\ \bibnamefont
  {McArdle}}, \bibinfo {author} {\bibfnamefont {Xiao}\ \bibnamefont {Yuan}}, \
  and\ \bibinfo {author} {\bibfnamefont {Simon}\ \bibnamefont {Benjamin}},\
  }\bibfield  {title} {\enquote {\bibinfo {title} {Error-mitigated digital
  quantum simulation},}\ }\href
  {https://journals.aps.org/prl/abstract/10.1103/PhysRevLett.122.180501}
  {\bibfield  {journal} {\bibinfo  {journal} {Phys. Rev. Lett.}\ }\textbf
  {\bibinfo {volume} {122}},\ \bibinfo {pages} {180501} (\bibinfo {year}
  {2019})}\BibitemShut {NoStop}%
\bibitem [{\citenamefont {Temme}\ \emph {et~al.}(2017)\citenamefont {Temme},
  \citenamefont {Bravyi},\ and\ \citenamefont {Gambetta}}]{temme2017error}%
  \BibitemOpen
  \bibfield  {author} {\bibinfo {author} {\bibfnamefont {Kristan}\ \bibnamefont
  {Temme}}, \bibinfo {author} {\bibfnamefont {Sergey}\ \bibnamefont {Bravyi}},
  \ and\ \bibinfo {author} {\bibfnamefont {Jay~M}\ \bibnamefont {Gambetta}},\
  }\bibfield  {title} {\enquote {\bibinfo {title} {Error mitigation for
  short-depth quantum circuits},}\ }\href
  {https://journals.aps.org/prl/abstract/10.1103/PhysRevLett.119.180509}
  {\bibfield  {journal} {\bibinfo  {journal} {Phys. Rev. Lett.}\ }\textbf
  {\bibinfo {volume} {119}},\ \bibinfo {pages} {180509} (\bibinfo {year}
  {2017})}\BibitemShut {NoStop}%
\bibitem [{\citenamefont {Sagastizabal}\ \emph {et~al.}(2019)\citenamefont
  {Sagastizabal}, \citenamefont {Bonet-Monroig}, \citenamefont {Singh},
  \citenamefont {Rol}, \citenamefont {Bultink}, \citenamefont {Fu},
  \citenamefont {Price}, \citenamefont {Ostroukh}, \citenamefont
  {Muthusubramanian}, \citenamefont {Bruno}, \citenamefont {Beekman},
  \citenamefont {Haider}, \citenamefont {O'Brien},\ and\ \citenamefont
  {DiCarlo}}]{sagastizabal2019error}%
  \BibitemOpen
  \bibfield  {author} {\bibinfo {author} {\bibfnamefont {R}~\bibnamefont
  {Sagastizabal}}, \bibinfo {author} {\bibfnamefont {X}~\bibnamefont
  {Bonet-Monroig}}, \bibinfo {author} {\bibfnamefont {M}~\bibnamefont {Singh}},
  \bibinfo {author} {\bibfnamefont {M~A}\ \bibnamefont {Rol}}, \bibinfo
  {author} {\bibfnamefont {C~C}\ \bibnamefont {Bultink}}, \bibinfo {author}
  {\bibfnamefont {X}~\bibnamefont {Fu}}, \bibinfo {author} {\bibfnamefont
  {C~H}\ \bibnamefont {Price}}, \bibinfo {author} {\bibfnamefont {V~P}\
  \bibnamefont {Ostroukh}}, \bibinfo {author} {\bibfnamefont {N}~\bibnamefont
  {Muthusubramanian}}, \bibinfo {author} {\bibfnamefont {A}~\bibnamefont
  {Bruno}}, \bibinfo {author} {\bibfnamefont {M}~\bibnamefont {Beekman}},
  \bibinfo {author} {\bibfnamefont {N}~\bibnamefont {Haider}}, \bibinfo
  {author} {\bibfnamefont {T~E}\ \bibnamefont {O'Brien}}, \ and\ \bibinfo
  {author} {\bibfnamefont {L}~\bibnamefont {DiCarlo}},\ }\bibfield  {title}
  {\enquote {\bibinfo {title} {Experimental error mitigation via symmetry
  verification in a variational quantum eigensolver},}\ }\href {\doibase
  10.1103/PhysRevA.100.010302} {\bibfield  {journal} {\bibinfo  {journal}
  {Phys. Rev. A}\ }\textbf {\bibinfo {volume} {100}},\ \bibinfo {pages}
  {010302} (\bibinfo {year} {2019})}\BibitemShut {NoStop}%
\bibitem [{\citenamefont {McClean}\ \emph
  {et~al.}(2017{\natexlab{a}})\citenamefont {McClean}, \citenamefont
  {Kimchi-Schwartz}, \citenamefont {Carter},\ and\ \citenamefont
  {de~Jong}}]{mcclean2017hybrid}%
  \BibitemOpen
  \bibfield  {author} {\bibinfo {author} {\bibfnamefont {Jarrod~R}\
  \bibnamefont {McClean}}, \bibinfo {author} {\bibfnamefont {Mollie~E}\
  \bibnamefont {Kimchi-Schwartz}}, \bibinfo {author} {\bibfnamefont {Jonathan}\
  \bibnamefont {Carter}}, \ and\ \bibinfo {author} {\bibfnamefont {Wibe~A}\
  \bibnamefont {de~Jong}},\ }\bibfield  {title} {\enquote {\bibinfo {title}
  {Hybrid quantum-classical hierarchy for mitigation of decoherence and
  determination of excited states},}\ }\href
  {https://journals.aps.org/pra/abstract/10.1103/PhysRevA.95.042308} {\bibfield
   {journal} {\bibinfo  {journal} {Phys. Rev. A}\ }\textbf {\bibinfo {volume}
  {95}},\ \bibinfo {pages} {042308} (\bibinfo {year}
  {2017}{\natexlab{a}})}\BibitemShut {NoStop}%
\bibitem [{\citenamefont {Otten}\ and\ \citenamefont
  {Gray}(2019)}]{otten2019accounting}%
  \BibitemOpen
  \bibfield  {author} {\bibinfo {author} {\bibfnamefont {Matthew}\ \bibnamefont
  {Otten}}\ and\ \bibinfo {author} {\bibfnamefont {Stephen~K}\ \bibnamefont
  {Gray}},\ }\bibfield  {title} {\enquote {\bibinfo {title} {Accounting for
  errors in quantum algorithms via individual error reduction},}\ }\href
  {https://www.nature.com/articles/s41534-019-0125-3} {\bibfield  {journal}
  {\bibinfo  {journal} {npj Quantum Inf.}\ }\textbf {\bibinfo {volume} {5}},\
  \bibinfo {pages} {11} (\bibinfo {year} {2019})}\BibitemShut {NoStop}%
\bibitem [{\citenamefont {O'Gorman}\ \emph {et~al.}(2019)\citenamefont
  {O'Gorman}, \citenamefont {Huggins}, \citenamefont {Rieffel},\ and\
  \citenamefont {Whaley}}]{o2019generalized}%
  \BibitemOpen
  \bibfield  {author} {\bibinfo {author} {\bibfnamefont {Bryan}\ \bibnamefont
  {O'Gorman}}, \bibinfo {author} {\bibfnamefont {William~J}\ \bibnamefont
  {Huggins}}, \bibinfo {author} {\bibfnamefont {Eleanor~G}\ \bibnamefont
  {Rieffel}}, \ and\ \bibinfo {author} {\bibfnamefont {K~Birgitta}\
  \bibnamefont {Whaley}},\ }\bibfield  {title} {\enquote {\bibinfo {title}
  {Generalized swap networks for near-term quantum computing},}\ }\href
  {https://arxiv.org/abs/1905.05118} {\bibfield  {journal} {\bibinfo  {journal}
  {arXiv:1905.05118}\ } (\bibinfo {year} {2019})}\BibitemShut {NoStop}%
\bibitem [{\citenamefont {Verteletskyi}\ \emph {et~al.}(2020)\citenamefont
  {Verteletskyi}, \citenamefont {Yen},\ and\ \citenamefont
  {Izmaylov}}]{1907.03358}%
  \BibitemOpen
  \bibfield  {author} {\bibinfo {author} {\bibfnamefont {Vladyslav}\
  \bibnamefont {Verteletskyi}}, \bibinfo {author} {\bibfnamefont {Tzu-Ching}\
  \bibnamefont {Yen}}, \ and\ \bibinfo {author} {\bibfnamefont {Artur~F}\
  \bibnamefont {Izmaylov}},\ }\bibfield  {title} {\enquote {\bibinfo {title}
  {Measurement optimization in the variational quantum eigensolver using a
  minimum clique cover},}\ }\href {\doibase 10.1063/1.5141458} {\bibfield
  {journal} {\bibinfo  {journal} {J. Chem. Phys.}\ }\textbf {\bibinfo {volume}
  {152}},\ \bibinfo {pages} {124114} (\bibinfo {year} {2020})}\BibitemShut
  {NoStop}%
\bibitem [{\citenamefont {Jena}\ \emph {et~al.}(2019)\citenamefont {Jena},
  \citenamefont {Genin},\ and\ \citenamefont {Mosca}}]{1907.07859}%
  \BibitemOpen
  \bibfield  {author} {\bibinfo {author} {\bibfnamefont {Andrew}\ \bibnamefont
  {Jena}}, \bibinfo {author} {\bibfnamefont {Scott}\ \bibnamefont {Genin}}, \
  and\ \bibinfo {author} {\bibfnamefont {Michele}\ \bibnamefont {Mosca}},\
  }\bibfield  {title} {\enquote {\bibinfo {title} {Pauli partitioning with
  respect to gate sets},}\ }\href {https://arxiv.org/abs/1907.07859} {\bibfield
   {journal} {\bibinfo  {journal} {arXiv:1907.07859}\ } (\bibinfo {year}
  {2019})}\BibitemShut {NoStop}%
\bibitem [{\citenamefont {Yen}\ \emph {et~al.}(2019)\citenamefont {Yen},
  \citenamefont {Verteletsky},\ and\ \citenamefont {Izmaylov}}]{1907.09386}%
  \BibitemOpen
  \bibfield  {author} {\bibinfo {author} {\bibfnamefont {Tzu-Ching}\
  \bibnamefont {Yen}}, \bibinfo {author} {\bibfnamefont {Vladyslav}\
  \bibnamefont {Verteletsky}}, \ and\ \bibinfo {author} {\bibfnamefont
  {Artur~F.}\ \bibnamefont {Izmaylov}},\ }\bibfield  {title} {\enquote
  {\bibinfo {title} {Measuring all compatible operators in one series of a
  single-qubit measurements using unitary transformations},}\ }\href
  {https://arxiv.org/abs/1907.09386} {\bibfield  {journal} {\bibinfo  {journal}
  {arXiv:1907.09386}\ } (\bibinfo {year} {2019})}\BibitemShut {NoStop}%
\bibitem [{\citenamefont {Izmaylov}\ \emph {et~al.}(2020)\citenamefont
  {Izmaylov}, \citenamefont {Yen}, \citenamefont {Lang},\ and\ \citenamefont
  {Verteletskyi}}]{1907.09040}%
  \BibitemOpen
  \bibfield  {author} {\bibinfo {author} {\bibfnamefont {Artur~F}\ \bibnamefont
  {Izmaylov}}, \bibinfo {author} {\bibfnamefont {Tzu-Ching}\ \bibnamefont
  {Yen}}, \bibinfo {author} {\bibfnamefont {Robert~A}\ \bibnamefont {Lang}}, \
  and\ \bibinfo {author} {\bibfnamefont {Vladyslav}\ \bibnamefont
  {Verteletskyi}},\ }\bibfield  {title} {\enquote {\bibinfo {title} {Unitary
  partitioning approach to the measurement problem in the variational quantum
  eigensolver method},}\ }\href {\doibase 10.1021/acs.jctc.9b00791} {\bibfield
  {journal} {\bibinfo  {journal} {J. Chem. Theory Comput.}\ }\textbf {\bibinfo
  {volume} {16}},\ \bibinfo {pages} {190--195} (\bibinfo {year}
  {2020})}\BibitemShut {NoStop}%
\bibitem [{\citenamefont {Izmaylov}\ \emph {et~al.}(2019)\citenamefont
  {Izmaylov}, \citenamefont {Yen},\ and\ \citenamefont
  {Ryabinkin}}]{izmaylov2019revising}%
  \BibitemOpen
  \bibfield  {author} {\bibinfo {author} {\bibfnamefont {Artur~F}\ \bibnamefont
  {Izmaylov}}, \bibinfo {author} {\bibfnamefont {Tzu-Ching}\ \bibnamefont
  {Yen}}, \ and\ \bibinfo {author} {\bibfnamefont {Ilya~G}\ \bibnamefont
  {Ryabinkin}},\ }\bibfield  {title} {\enquote {\bibinfo {title} {Revising the
  measurement process in the variational quantum eigensolver: is it possible to
  reduce the number of separately measured operators?}}\ }\href
  {https://pubs.rsc.org/en/content/articlelanding/2019/sc/c8sc05592k}
  {\bibfield  {journal} {\bibinfo  {journal} {Chem. Sci.}\ }\textbf {\bibinfo
  {volume} {10}},\ \bibinfo {pages} {3746--3755} (\bibinfo {year}
  {2019})}\BibitemShut {NoStop}%
\bibitem [{\citenamefont {Gokhale}\ \emph {et~al.}(2019)\citenamefont
  {Gokhale}, \citenamefont {Angiuli}, \citenamefont {Ding}, \citenamefont
  {Gui}, \citenamefont {Tomesh}, \citenamefont {Suchara}, \citenamefont
  {Martonosi},\ and\ \citenamefont {Chong}}]{Gokhale2019-cv}%
  \BibitemOpen
  \bibfield  {author} {\bibinfo {author} {\bibfnamefont {Pranav}\ \bibnamefont
  {Gokhale}}, \bibinfo {author} {\bibfnamefont {Olivia}\ \bibnamefont
  {Angiuli}}, \bibinfo {author} {\bibfnamefont {Yongshan}\ \bibnamefont
  {Ding}}, \bibinfo {author} {\bibfnamefont {Kaiwen}\ \bibnamefont {Gui}},
  \bibinfo {author} {\bibfnamefont {Teague}\ \bibnamefont {Tomesh}}, \bibinfo
  {author} {\bibfnamefont {Martin}\ \bibnamefont {Suchara}}, \bibinfo {author}
  {\bibfnamefont {Margaret}\ \bibnamefont {Martonosi}}, \ and\ \bibinfo
  {author} {\bibfnamefont {Frederic~T}\ \bibnamefont {Chong}},\ }\bibfield
  {title} {\enquote {\bibinfo {title} {Minimizing state preparations in
  variational quantum eigensolver by partitioning into commuting families},}\
  }\href {https://arxiv.org/abs/1907.13623} {\bibfield  {journal} {\bibinfo
  {journal} {arXiv:1907.13623}\ } (\bibinfo {year} {2019})}\BibitemShut
  {NoStop}%
\bibitem [{\citenamefont {Rubin}\ \emph {et~al.}(2018)\citenamefont {Rubin},
  \citenamefont {Babbush},\ and\ \citenamefont
  {McClean}}]{rubin2018application}%
  \BibitemOpen
  \bibfield  {author} {\bibinfo {author} {\bibfnamefont {Nicholas~C}\
  \bibnamefont {Rubin}}, \bibinfo {author} {\bibfnamefont {Ryan}\ \bibnamefont
  {Babbush}}, \ and\ \bibinfo {author} {\bibfnamefont {Jarrod}\ \bibnamefont
  {McClean}},\ }\bibfield  {title} {\enquote {\bibinfo {title} {Application of
  fermionic marginal constraints to hybrid quantum algorithms},}\ }\href
  {https://iopscience.iop.org/article/10.1088/1367-2630/aab919/meta} {\bibfield
   {journal} {\bibinfo  {journal} {New J. Phys.}\ }\textbf {\bibinfo {volume}
  {20}},\ \bibinfo {pages} {053020} (\bibinfo {year} {2018})}\BibitemShut
  {NoStop}%
\bibitem [{\citenamefont {Setia}\ and\ \citenamefont
  {Whitfield}(2018)}]{setia2018bravyi}%
  \BibitemOpen
  \bibfield  {author} {\bibinfo {author} {\bibfnamefont {Kanav}\ \bibnamefont
  {Setia}}\ and\ \bibinfo {author} {\bibfnamefont {James~D}\ \bibnamefont
  {Whitfield}},\ }\bibfield  {title} {\enquote {\bibinfo {title} {Bravyi-kitaev
  superfast simulation of electronic structure on a quantum computer},}\ }\href
  {https://aip.scitation.org/doi/10.1063/1.5019371} {\bibfield  {journal}
  {\bibinfo  {journal} {J. Chem. Phys.}\ }\textbf {\bibinfo {volume} {148}},\
  \bibinfo {pages} {164104} (\bibinfo {year} {2018})}\BibitemShut {NoStop}%
\bibitem [{\citenamefont {Jiang}\ \emph {et~al.}(2018)\citenamefont {Jiang},
  \citenamefont {McClean}, \citenamefont {Babbush},\ and\ \citenamefont
  {Neven}}]{jiang2018majorana}%
  \BibitemOpen
  \bibfield  {author} {\bibinfo {author} {\bibfnamefont {Zhang}\ \bibnamefont
  {Jiang}}, \bibinfo {author} {\bibfnamefont {Jarrod}\ \bibnamefont {McClean}},
  \bibinfo {author} {\bibfnamefont {Ryan}\ \bibnamefont {Babbush}}, \ and\
  \bibinfo {author} {\bibfnamefont {Hartmut}\ \bibnamefont {Neven}},\
  }\bibfield  {title} {\enquote {\bibinfo {title} {Majorana loop stabilizer
  codes for error correction of fermionic quantum simulations},}\ }\href
  {https://arxiv.org/abs/1812.08190} {\bibfield  {journal} {\bibinfo  {journal}
  {arXiv:1812.08190}\ } (\bibinfo {year} {2018})}\BibitemShut {NoStop}%
\bibitem [{\citenamefont {Bravyi}\ and\ \citenamefont
  {Kitaev}(2002)}]{bravyi2002fermionic}%
  \BibitemOpen
  \bibfield  {author} {\bibinfo {author} {\bibfnamefont {Sergey~B}\
  \bibnamefont {Bravyi}}\ and\ \bibinfo {author} {\bibfnamefont {Alexei~Yu}\
  \bibnamefont {Kitaev}},\ }\bibfield  {title} {\enquote {\bibinfo {title}
  {Fermionic quantum computation},}\ }\href
  {https://www.sciencedirect.com/science/article/pii/S0003491602962548}
  {\bibfield  {journal} {\bibinfo  {journal} {Ann. Phys.}\ }\textbf {\bibinfo
  {volume} {298}},\ \bibinfo {pages} {210--226} (\bibinfo {year}
  {2002})}\BibitemShut {NoStop}%
\bibitem [{\citenamefont {Babbush}\ \emph {et~al.}(2018)\citenamefont
  {Babbush}, \citenamefont {Wiebe}, \citenamefont {McClean}, \citenamefont
  {McClain}, \citenamefont {Neven},\ and\ \citenamefont
  {Chan}}]{babbush2018low}%
  \BibitemOpen
  \bibfield  {author} {\bibinfo {author} {\bibfnamefont {Ryan}\ \bibnamefont
  {Babbush}}, \bibinfo {author} {\bibfnamefont {Nathan}\ \bibnamefont {Wiebe}},
  \bibinfo {author} {\bibfnamefont {Jarrod}\ \bibnamefont {McClean}}, \bibinfo
  {author} {\bibfnamefont {James}\ \bibnamefont {McClain}}, \bibinfo {author}
  {\bibfnamefont {Hartmut}\ \bibnamefont {Neven}}, \ and\ \bibinfo {author}
  {\bibfnamefont {Garnet Kin-Lic}\ \bibnamefont {Chan}},\ }\bibfield  {title}
  {\enquote {\bibinfo {title} {{Low-Depth Quantum Simulation of Materials}},}\
  }\href {https://journals.aps.org/prx/abstract/10.1103/PhysRevX.8.011044}
  {\bibfield  {journal} {\bibinfo  {journal} {Physical Review X}\ }\textbf
  {\bibinfo {volume} {8}},\ \bibinfo {pages} {011044} (\bibinfo {year}
  {2018})}\BibitemShut {NoStop}%
\bibitem [{\citenamefont {Kivlichan}\ \emph {et~al.}(2018)\citenamefont
  {Kivlichan}, \citenamefont {McClean}, \citenamefont {Wiebe}, \citenamefont
  {Gidney}, \citenamefont {Aspuru-Guzik}, \citenamefont {Chan},\ and\
  \citenamefont {Babbush}}]{kivlichan2018quantum}%
  \BibitemOpen
  \bibfield  {author} {\bibinfo {author} {\bibfnamefont {Ian~D}\ \bibnamefont
  {Kivlichan}}, \bibinfo {author} {\bibfnamefont {Jarrod}\ \bibnamefont
  {McClean}}, \bibinfo {author} {\bibfnamefont {Nathan}\ \bibnamefont {Wiebe}},
  \bibinfo {author} {\bibfnamefont {Craig}\ \bibnamefont {Gidney}}, \bibinfo
  {author} {\bibfnamefont {Al{\'a}n}\ \bibnamefont {Aspuru-Guzik}}, \bibinfo
  {author} {\bibfnamefont {Garnet Kin-Lic}\ \bibnamefont {Chan}}, \ and\
  \bibinfo {author} {\bibfnamefont {Ryan}\ \bibnamefont {Babbush}},\ }\bibfield
   {title} {\enquote {\bibinfo {title} {Quantum simulation of electronic
  structure with linear depth and connectivity},}\ }\href
  {https://link.aps.org/doi/10.1103/PhysRevLett.120.110501} {\bibfield
  {journal} {\bibinfo  {journal} {Phys. Rev. Lett.}\ }\textbf {\bibinfo
  {volume} {120}},\ \bibinfo {pages} {110501} (\bibinfo {year}
  {2018})}\BibitemShut {NoStop}%
\bibitem [{\citenamefont {Motta}\ \emph {et~al.}(2018)\citenamefont {Motta},
  \citenamefont {Ye}, \citenamefont {McClean}, \citenamefont {Li},
  \citenamefont {Minnich}, \citenamefont {Babbush},\ and\ \citenamefont
  {Chan}}]{motta2018low}%
  \BibitemOpen
  \bibfield  {author} {\bibinfo {author} {\bibfnamefont {Mario}\ \bibnamefont
  {Motta}}, \bibinfo {author} {\bibfnamefont {Erika}\ \bibnamefont {Ye}},
  \bibinfo {author} {\bibfnamefont {Jarrod~R}\ \bibnamefont {McClean}},
  \bibinfo {author} {\bibfnamefont {Zhendong}\ \bibnamefont {Li}}, \bibinfo
  {author} {\bibfnamefont {Austin~J}\ \bibnamefont {Minnich}}, \bibinfo
  {author} {\bibfnamefont {Ryan}\ \bibnamefont {Babbush}}, \ and\ \bibinfo
  {author} {\bibfnamefont {Garnet~Kin}\ \bibnamefont {Chan}},\ }\bibfield
  {title} {\enquote {\bibinfo {title} {Low rank representations for quantum
  simulation of electronic structure},}\ }\href
  {https://arxiv.org/abs/1808.02625} {\bibfield  {journal} {\bibinfo  {journal}
  {arXiv:1808.02625}\ } (\bibinfo {year} {2018})}\BibitemShut {NoStop}%
\bibitem [{\citenamefont {Poulin}\ \emph {et~al.}(2015)\citenamefont {Poulin},
  \citenamefont {Hastings}, \citenamefont {Wecker}, \citenamefont {Wiebe},
  \citenamefont {Doherty},\ and\ \citenamefont {Troyer}}]{poulin2014trotter}%
  \BibitemOpen
  \bibfield  {author} {\bibinfo {author} {\bibfnamefont {David}\ \bibnamefont
  {Poulin}}, \bibinfo {author} {\bibfnamefont {M~B}\ \bibnamefont {Hastings}},
  \bibinfo {author} {\bibfnamefont {Dave}\ \bibnamefont {Wecker}}, \bibinfo
  {author} {\bibfnamefont {Nathan}\ \bibnamefont {Wiebe}}, \bibinfo {author}
  {\bibfnamefont {Andrew~C}\ \bibnamefont {Doherty}}, \ and\ \bibinfo {author}
  {\bibfnamefont {Matthias}\ \bibnamefont {Troyer}},\ }\bibfield  {title}
  {\enquote {\bibinfo {title} {{The Trotter Step Size Required for Accurate
  Quantum Simulation of Quantum Chemistry}},}\ }\href
  {http://arxiv.org/abs/1406.4920} {\bibfield  {journal} {\bibinfo  {journal}
  {Quantum Information {\&} Computation}\ }\textbf {\bibinfo {volume} {15}},\
  \bibinfo {pages} {361--384} (\bibinfo {year} {2015})}\BibitemShut {NoStop}%
\bibitem [{\citenamefont {Berry}\ \emph {et~al.}(2019)\citenamefont {Berry},
  \citenamefont {Gidney}, \citenamefont {Motta}, \citenamefont {McClean},\ and\
  \citenamefont {Babbush}}]{berry2019qubitization}%
  \BibitemOpen
  \bibfield  {author} {\bibinfo {author} {\bibfnamefont {Dominic~W}\
  \bibnamefont {Berry}}, \bibinfo {author} {\bibfnamefont {Craig}\ \bibnamefont
  {Gidney}}, \bibinfo {author} {\bibfnamefont {Mario}\ \bibnamefont {Motta}},
  \bibinfo {author} {\bibfnamefont {Jarrod~R}\ \bibnamefont {McClean}}, \ and\
  \bibinfo {author} {\bibfnamefont {Ryan}\ \bibnamefont {Babbush}},\ }\bibfield
   {title} {\enquote {\bibinfo {title} {Qubitization of arbitrary basis quantum
  chemistry leveraging sparsity and low rank factorization},}\ }\href {\doibase
  10.22331/q-2019-12-02-208} {\bibfield  {journal} {\bibinfo  {journal}
  {Quantum}\ }\textbf {\bibinfo {volume} {3}},\ \bibinfo {pages} {208}
  (\bibinfo {year} {2019})}\BibitemShut {NoStop}%
\bibitem [{\citenamefont {Whitten}(1973)}]{whitten1973coulombic}%
  \BibitemOpen
  \bibfield  {author} {\bibinfo {author} {\bibfnamefont {Jerry~L}\ \bibnamefont
  {Whitten}},\ }\bibfield  {title} {\enquote {\bibinfo {title} {Coulombic
  potential energy integrals and approximations},}\ }\href
  {https://aip.scitation.org/doi/abs/10.1063/1.1679012} {\bibfield  {journal}
  {\bibinfo  {journal} {J. Chem. Phys.}\ }\textbf {\bibinfo {volume} {58}},\
  \bibinfo {pages} {4496--4501} (\bibinfo {year} {1973})}\BibitemShut {NoStop}%
\bibitem [{\citenamefont {Aquilante}\ \emph {et~al.}(2010)\citenamefont
  {Aquilante}, \citenamefont {De~Vico}, \citenamefont {Ferr{\'e}},
  \citenamefont {Ghigo}, \citenamefont {Malmqvist}, \citenamefont
  {Neogr{\'a}dy}, \citenamefont {Pedersen}, \citenamefont {Pito{\v{n}}{\'a}k},
  \citenamefont {Reiher}, \citenamefont {Roos} \emph
  {et~al.}}]{aquilante2010molcas}%
  \BibitemOpen
  \bibfield  {author} {\bibinfo {author} {\bibfnamefont {Francesco}\
  \bibnamefont {Aquilante}}, \bibinfo {author} {\bibfnamefont {Luca}\
  \bibnamefont {De~Vico}}, \bibinfo {author} {\bibfnamefont {Nicolas}\
  \bibnamefont {Ferr{\'e}}}, \bibinfo {author} {\bibfnamefont {Giovanni}\
  \bibnamefont {Ghigo}}, \bibinfo {author} {\bibfnamefont {Per-{\aa}ke}\
  \bibnamefont {Malmqvist}}, \bibinfo {author} {\bibfnamefont {Pavel}\
  \bibnamefont {Neogr{\'a}dy}}, \bibinfo {author} {\bibfnamefont
  {Thomas~Bondo}\ \bibnamefont {Pedersen}}, \bibinfo {author} {\bibfnamefont
  {Michal}\ \bibnamefont {Pito{\v{n}}{\'a}k}}, \bibinfo {author} {\bibfnamefont
  {Markus}\ \bibnamefont {Reiher}}, \bibinfo {author} {\bibfnamefont
  {Bj{\"o}rn~O}\ \bibnamefont {Roos}},  \emph {et~al.},\ }\bibfield  {title}
  {\enquote {\bibinfo {title} {Molcas 7: the next generation},}\ }\href
  {https://onlinelibrary.wiley.com/doi/full/10.1002/jcc.21318} {\bibfield
  {journal} {\bibinfo  {journal} {J. Comput. Chem.}\ }\textbf {\bibinfo
  {volume} {31}},\ \bibinfo {pages} {224--247} (\bibinfo {year}
  {2010})}\BibitemShut {NoStop}%
\bibitem [{\citenamefont {Pedersen}\ \emph {et~al.}(2009)\citenamefont
  {Pedersen}, \citenamefont {Aquilante},\ and\ \citenamefont
  {Lindh}}]{Pedersen2009-xc}%
  \BibitemOpen
  \bibfield  {author} {\bibinfo {author} {\bibfnamefont {Thomas~Bondo}\
  \bibnamefont {Pedersen}}, \bibinfo {author} {\bibfnamefont {Francesco}\
  \bibnamefont {Aquilante}}, \ and\ \bibinfo {author} {\bibfnamefont {Roland}\
  \bibnamefont {Lindh}},\ }\bibfield  {title} {\enquote {\bibinfo {title}
  {Density fitting with auxiliary basis sets from cholesky decompositions},}\
  }\href {\doibase 10.1007/s00214-009-0608-y} {\bibfield  {journal} {\bibinfo
  {journal} {Theor. Chem. Acc.}\ }\textbf {\bibinfo {volume} {124}},\ \bibinfo
  {pages} {1--10} (\bibinfo {year} {2009})}\BibitemShut {NoStop}%
\bibitem [{\citenamefont {Beebe}\ and\ \citenamefont
  {Linderberg}(1977)}]{beebe1977simplifications}%
  \BibitemOpen
  \bibfield  {author} {\bibinfo {author} {\bibfnamefont {Nelson~HF}\
  \bibnamefont {Beebe}}\ and\ \bibinfo {author} {\bibfnamefont {Jan}\
  \bibnamefont {Linderberg}},\ }\bibfield  {title} {\enquote {\bibinfo {title}
  {Simplifications in the generation and transformation of two-electron
  integrals in molecular calculations},}\ }\href
  {https://onlinelibrary.wiley.com/doi/abs/10.1002/qua.560120408} {\bibfield
  {journal} {\bibinfo  {journal} {Int. J. Quantum Chem.}\ }\textbf {\bibinfo
  {volume} {12}},\ \bibinfo {pages} {683--705} (\bibinfo {year}
  {1977})}\BibitemShut {NoStop}%
\bibitem [{\citenamefont {Koch}\ \emph {et~al.}(2003)\citenamefont {Koch},
  \citenamefont {S{\'a}nchez~de Mer{\'a}s},\ and\ \citenamefont
  {Pedersen}}]{koch2003reduced}%
  \BibitemOpen
  \bibfield  {author} {\bibinfo {author} {\bibfnamefont {Henrik}\ \bibnamefont
  {Koch}}, \bibinfo {author} {\bibfnamefont {Alfredo}\ \bibnamefont
  {S{\'a}nchez~de Mer{\'a}s}}, \ and\ \bibinfo {author} {\bibfnamefont
  {Thomas~Bondo}\ \bibnamefont {Pedersen}},\ }\bibfield  {title} {\enquote
  {\bibinfo {title} {Reduced scaling in electronic structure calculations using
  cholesky decompositions},}\ }\href
  {https://aip.scitation.org/doi/abs/10.1063/1.1578621} {\bibfield  {journal}
  {\bibinfo  {journal} {J. Chem. Phys.}\ }\textbf {\bibinfo {volume} {118}},\
  \bibinfo {pages} {9481--9484} (\bibinfo {year} {2003})}\BibitemShut {NoStop}%
\bibitem [{\citenamefont {Purwanto}\ \emph {et~al.}(2011)\citenamefont
  {Purwanto}, \citenamefont {Krakauer}, \citenamefont {Virgus},\ and\
  \citenamefont {Zhang}}]{purwanto2011assessing}%
  \BibitemOpen
  \bibfield  {author} {\bibinfo {author} {\bibfnamefont {Wirawan}\ \bibnamefont
  {Purwanto}}, \bibinfo {author} {\bibfnamefont {Henry}\ \bibnamefont
  {Krakauer}}, \bibinfo {author} {\bibfnamefont {Yudistira}\ \bibnamefont
  {Virgus}}, \ and\ \bibinfo {author} {\bibfnamefont {Shiwei}\ \bibnamefont
  {Zhang}},\ }\bibfield  {title} {\enquote {\bibinfo {title} {Assessing weak
  hydrogen binding on ca+ centers: An accurate many-body study with large basis
  sets},}\ }\href {https://aip.scitation.org/doi/full/10.1063/1.3654002}
  {\bibfield  {journal} {\bibinfo  {journal} {J. Chem. Phys.}\ }\textbf
  {\bibinfo {volume} {135}},\ \bibinfo {pages} {164105} (\bibinfo {year}
  {2011})}\BibitemShut {NoStop}%
\bibitem [{\citenamefont {Mardirossian}\ \emph {et~al.}(2018)\citenamefont
  {Mardirossian}, \citenamefont {McClain},\ and\ \citenamefont
  {Chan}}]{mardirossian2018lowering}%
  \BibitemOpen
  \bibfield  {author} {\bibinfo {author} {\bibfnamefont {Narbe}\ \bibnamefont
  {Mardirossian}}, \bibinfo {author} {\bibfnamefont {James~D}\ \bibnamefont
  {McClain}}, \ and\ \bibinfo {author} {\bibfnamefont {Garnet Kin-Lic}\
  \bibnamefont {Chan}},\ }\bibfield  {title} {\enquote {\bibinfo {title}
  {Lowering of the complexity of quantum chemistry methods by choice of
  representation},}\ }\href {https://aip.scitation.org/doi/10.1063/1.5007779}
  {\bibfield  {journal} {\bibinfo  {journal} {J. Chem. Phys.}\ }\textbf
  {\bibinfo {volume} {148}},\ \bibinfo {pages} {044106} (\bibinfo {year}
  {2018})}\BibitemShut {NoStop}%
\bibitem [{\citenamefont {Peng}\ and\ \citenamefont
  {Kowalski}(2017)}]{Peng2017-oo}%
  \BibitemOpen
  \bibfield  {author} {\bibinfo {author} {\bibfnamefont {Bo}~\bibnamefont
  {Peng}}\ and\ \bibinfo {author} {\bibfnamefont {Karol}\ \bibnamefont
  {Kowalski}},\ }\bibfield  {title} {\enquote {\bibinfo {title} {Highly
  efficient and scalable compound decomposition of {Two-Electron} integral
  tensor and its application in coupled cluster calculations},}\ }\href
  {\doibase 10.1021/acs.jctc.7b00605} {\bibfield  {journal} {\bibinfo
  {journal} {J. Chem. Theory Comput.}\ }\textbf {\bibinfo {volume} {13}},\
  \bibinfo {pages} {4179--4192} (\bibinfo {year} {2017})}\BibitemShut {NoStop}%
\bibitem [{\citenamefont {R{\o}eggen}\ and\ \citenamefont
  {Wisl{\o}ff-Nilssen}(1986)}]{Roeggen1986-ib}%
  \BibitemOpen
  \bibfield  {author} {\bibinfo {author} {\bibfnamefont {I}~\bibnamefont
  {R{\o}eggen}}\ and\ \bibinfo {author} {\bibfnamefont {E}~\bibnamefont
  {Wisl{\o}ff-Nilssen}},\ }\bibfield  {title} {\enquote {\bibinfo {title} {On
  the {Beebe-Linderberg} two-electron integral approximation},}\ }\href
  {\doibase 10.1016/0009-2614(86)80099-9} {\bibfield  {journal} {\bibinfo
  {journal} {Chem. Phys. Lett.}\ }\textbf {\bibinfo {volume} {132}},\ \bibinfo
  {pages} {154--160} (\bibinfo {year} {1986})}\BibitemShut {NoStop}%
\bibitem [{\citenamefont {R{\o}eggen}\ and\ \citenamefont
  {Johansen}(2008)}]{Roeggen2008-bt}%
  \BibitemOpen
  \bibfield  {author} {\bibinfo {author} {\bibfnamefont {I}~\bibnamefont
  {R{\o}eggen}}\ and\ \bibinfo {author} {\bibfnamefont {Tor}\ \bibnamefont
  {Johansen}},\ }\bibfield  {title} {\enquote {\bibinfo {title} {Cholesky
  decomposition of the two-electron integral matrix in electronic structure
  calculations},}\ }\href {\doibase 10.1063/1.2925269} {\bibfield  {journal}
  {\bibinfo  {journal} {J. Chem. Phys.}\ }\textbf {\bibinfo {volume} {128}},\
  \bibinfo {pages} {194107} (\bibinfo {year} {2008})}\BibitemShut {NoStop}%
\bibitem [{\citenamefont {Boman}\ \emph {et~al.}(2008)\citenamefont {Boman},
  \citenamefont {Koch},\ and\ \citenamefont {S{\'a}nchez~de
  Mer{\'a}s}}]{Boman2008-mm}%
  \BibitemOpen
  \bibfield  {author} {\bibinfo {author} {\bibfnamefont {Linus}\ \bibnamefont
  {Boman}}, \bibinfo {author} {\bibfnamefont {Henrik}\ \bibnamefont {Koch}}, \
  and\ \bibinfo {author} {\bibfnamefont {Alfredo}\ \bibnamefont {S{\'a}nchez~de
  Mer{\'a}s}},\ }\bibfield  {title} {\enquote {\bibinfo {title} {Method
  specific cholesky decomposition: coulomb and exchange energies},}\ }\href
  {\doibase 10.1063/1.2988315} {\bibfield  {journal} {\bibinfo  {journal} {J.
  Chem. Phys.}\ }\textbf {\bibinfo {volume} {129}},\ \bibinfo {pages} {134107}
  (\bibinfo {year} {2008})}\BibitemShut {NoStop}%
\bibitem [{\citenamefont {Clements}\ \emph {et~al.}(2016)\citenamefont
  {Clements}, \citenamefont {Humphreys}, \citenamefont {Metcalf}, \citenamefont
  {Kolthammer},\ and\ \citenamefont {Walmsley}}]{clements2016optimal}%
  \BibitemOpen
  \bibfield  {author} {\bibinfo {author} {\bibfnamefont {William~R}\
  \bibnamefont {Clements}}, \bibinfo {author} {\bibfnamefont {Peter~C}\
  \bibnamefont {Humphreys}}, \bibinfo {author} {\bibfnamefont {Benjamin~J}\
  \bibnamefont {Metcalf}}, \bibinfo {author} {\bibfnamefont {W~Steven}\
  \bibnamefont {Kolthammer}}, \ and\ \bibinfo {author} {\bibfnamefont {Ian~A}\
  \bibnamefont {Walmsley}},\ }\bibfield  {title} {\enquote {\bibinfo {title}
  {Optimal design for universal multiport interferometers},}\ }\href
  {https://www.osapublishing.org/optica/abstract.cfm?uri=optica-3-12-1460}
  {\bibfield  {journal} {\bibinfo  {journal} {Optica}\ }\textbf {\bibinfo
  {volume} {3}},\ \bibinfo {pages} {1460--1465} (\bibinfo {year}
  {2016})}\BibitemShut {NoStop}%
\bibitem [{\citenamefont {McClean}\ \emph
  {et~al.}(2017{\natexlab{b}})\citenamefont {McClean}, \citenamefont
  {Kivlichan}, \citenamefont {Sung}, \citenamefont {Steiger}, \citenamefont
  {Cao}, \citenamefont {Dai}, \citenamefont {Fried}, \citenamefont {Gidney},
  \citenamefont {Gimby}, \citenamefont {Gokhale} \emph
  {et~al.}}]{mcclean2017openfermion}%
  \BibitemOpen
  \bibfield  {author} {\bibinfo {author} {\bibfnamefont {Jarrod~R}\
  \bibnamefont {McClean}}, \bibinfo {author} {\bibfnamefont {Ian~D}\
  \bibnamefont {Kivlichan}}, \bibinfo {author} {\bibfnamefont {Kevin~J}\
  \bibnamefont {Sung}}, \bibinfo {author} {\bibfnamefont {Damian~S}\
  \bibnamefont {Steiger}}, \bibinfo {author} {\bibfnamefont {Yudong}\
  \bibnamefont {Cao}}, \bibinfo {author} {\bibfnamefont {Chengyu}\ \bibnamefont
  {Dai}}, \bibinfo {author} {\bibfnamefont {E~Schuyler}\ \bibnamefont {Fried}},
  \bibinfo {author} {\bibfnamefont {Craig}\ \bibnamefont {Gidney}}, \bibinfo
  {author} {\bibfnamefont {Brendan}\ \bibnamefont {Gimby}}, \bibinfo {author}
  {\bibfnamefont {Pranav}\ \bibnamefont {Gokhale}},  \emph {et~al.},\
  }\bibfield  {title} {\enquote {\bibinfo {title} {Openfermion: the electronic
  structure package for quantum computers},}\ }\href
  {https://arxiv.org/abs/1710.07629} {\bibfield  {journal} {\bibinfo  {journal}
  {arXiv:1710.07629}\ } (\bibinfo {year} {2017}{\natexlab{b}})}\BibitemShut
  {NoStop}%
\bibitem [{\citenamefont {Parrish}\ \emph {et~al.}(2017)\citenamefont
  {Parrish}, \citenamefont {Burns}, \citenamefont {Smith}, \citenamefont
  {Simmonett}, \citenamefont {DePrince~III}, \citenamefont {Hohenstein},
  \citenamefont {Bozkaya}, \citenamefont {Sokolov}, \citenamefont {Di~Remigio},
  \citenamefont {Richard} \emph {et~al.}}]{parrish2017psi4}%
  \BibitemOpen
  \bibfield  {author} {\bibinfo {author} {\bibfnamefont {Robert~M}\
  \bibnamefont {Parrish}}, \bibinfo {author} {\bibfnamefont {Lori~A}\
  \bibnamefont {Burns}}, \bibinfo {author} {\bibfnamefont {Daniel~GA}\
  \bibnamefont {Smith}}, \bibinfo {author} {\bibfnamefont {Andrew~C}\
  \bibnamefont {Simmonett}}, \bibinfo {author} {\bibfnamefont {A~Eugene}\
  \bibnamefont {DePrince~III}}, \bibinfo {author} {\bibfnamefont {Edward~G}\
  \bibnamefont {Hohenstein}}, \bibinfo {author} {\bibfnamefont {Ugur}\
  \bibnamefont {Bozkaya}}, \bibinfo {author} {\bibfnamefont {Alexander~Yu}\
  \bibnamefont {Sokolov}}, \bibinfo {author} {\bibfnamefont {Roberto}\
  \bibnamefont {Di~Remigio}}, \bibinfo {author} {\bibfnamefont {Ryan~M}\
  \bibnamefont {Richard}},  \emph {et~al.},\ }\bibfield  {title} {\enquote
  {\bibinfo {title} {Psi4 1.1: An open-source electronic structure program
  emphasizing automation, advanced libraries, and interoperability},}\ }\href
  {https://pubs.acs.org/doi/10.1021/acs.jctc.7b00174} {\bibfield  {journal}
  {\bibinfo  {journal} {J. Chem. Theory Comput.}\ }\textbf {\bibinfo {volume}
  {13}},\ \bibinfo {pages} {3185--3197} (\bibinfo {year} {2017})}\BibitemShut
  {NoStop}%
\bibitem [{\citenamefont {Huggins}\ \emph {et~al.}(2019)\citenamefont
  {Huggins}, \citenamefont {Lee}, \citenamefont {Baek}, \citenamefont
  {O'Gorman},\ and\ \citenamefont {Birgitta~Whaley}}]{Huggins2019-gi}%
  \BibitemOpen
  \bibfield  {author} {\bibinfo {author} {\bibfnamefont {William~J}\
  \bibnamefont {Huggins}}, \bibinfo {author} {\bibfnamefont {Joonho}\
  \bibnamefont {Lee}}, \bibinfo {author} {\bibfnamefont {Unpil}\ \bibnamefont
  {Baek}}, \bibinfo {author} {\bibfnamefont {Bryan}\ \bibnamefont {O'Gorman}},
  \ and\ \bibinfo {author} {\bibfnamefont {K}~\bibnamefont {Birgitta~Whaley}},\
  }\bibfield  {title} {\enquote {\bibinfo {title} {A {Non-Orthogonal}
  variational quantum eigensolver},}\ }\href {http://arxiv.org/abs/1909.09114}
  {\bibfield  {journal} {\bibinfo  {journal} {arXiv:1909.09114}\ } (\bibinfo
  {year} {2019})}\BibitemShut {NoStop}%
\bibitem [{Note1()}]{Note1}%
  \BibitemOpen
  \bibinfo {note} {In the process of preparing this manuscript we have become
  aware of several recent works that employ more sophisticated strategies for
  grouping Pauli words together or employing a different family of unitary
  transformations than those we consider to enhance the measurement
  process~\cite {1907.03358, 1907.07859, 1907.09386, 1907.09040}. It would be
  an interesting subject of future work to calculate and compare the number of
  circuit repetitions required by these approaches.}\BibitemShut {Stop}%
\bibitem [{\citenamefont {on~the Many-Electron~Problem}\ \emph
  {et~al.}(2017)\citenamefont {on~the Many-Electron~Problem}, \citenamefont
  {Motta}, \citenamefont {Ceperley}, \citenamefont {Chan}, \citenamefont
  {Gomez}, \citenamefont {Gull}, \citenamefont {Guo}, \citenamefont
  {Jiménez-Hoyos}, \citenamefont {Lan}, \citenamefont {Li},\ and\
  \citenamefont {et~al.}}]{Simons_collab_2017}%
  \BibitemOpen
  \bibfield  {author} {\bibinfo {author} {\bibfnamefont {Simons~Collaboration}\
  \bibnamefont {on~the Many-Electron~Problem}}, \bibinfo {author}
  {\bibfnamefont {Mario}\ \bibnamefont {Motta}}, \bibinfo {author}
  {\bibfnamefont {David~M.}\ \bibnamefont {Ceperley}}, \bibinfo {author}
  {\bibfnamefont {Garnet Kin-Lic}\ \bibnamefont {Chan}}, \bibinfo {author}
  {\bibfnamefont {John~A.}\ \bibnamefont {Gomez}}, \bibinfo {author}
  {\bibfnamefont {Emanuel}\ \bibnamefont {Gull}}, \bibinfo {author}
  {\bibfnamefont {Sheng}\ \bibnamefont {Guo}}, \bibinfo {author} {\bibfnamefont
  {Carlos~A.}\ \bibnamefont {Jiménez-Hoyos}}, \bibinfo {author} {\bibfnamefont
  {Tran~Nguyen}\ \bibnamefont {Lan}}, \bibinfo {author} {\bibfnamefont {Jia}\
  \bibnamefont {Li}}, \ and\ \bibinfo {author} {\bibnamefont {et~al.}},\
  }\bibfield  {title} {\enquote {\bibinfo {title} {Towards the solution of the
  many-electron problem in real materials: Equation of state of the hydrogen
  chain with state-of-the-art many-body methods},}\ }\href {\doibase
  10.1103/PhysRevX.7.031059} {\bibfield  {journal} {\bibinfo  {journal} {Phys,
  Rev. X}\ }\textbf {\bibinfo {volume} {7}},\ \bibinfo {pages} {031059}
  (\bibinfo {year} {2017})}\BibitemShut {NoStop}%
\bibitem [{\citenamefont {Granade}\ \emph {et~al.}(2012)\citenamefont
  {Granade}, \citenamefont {Ferrie}, \citenamefont {Wiebe},\ and\ \citenamefont
  {Cory}}]{granade2012robust}%
  \BibitemOpen
  \bibfield  {author} {\bibinfo {author} {\bibfnamefont {Christopher~E}\
  \bibnamefont {Granade}}, \bibinfo {author} {\bibfnamefont {Christopher}\
  \bibnamefont {Ferrie}}, \bibinfo {author} {\bibfnamefont {Nathan}\
  \bibnamefont {Wiebe}}, \ and\ \bibinfo {author} {\bibfnamefont {David~G}\
  \bibnamefont {Cory}},\ }\bibfield  {title} {\enquote {\bibinfo {title}
  {Robust online hamiltonian learning},}\ }\href@noop {} {\bibfield  {journal}
  {\bibinfo  {journal} {New Journal of Physics}\ }\textbf {\bibinfo {volume}
  {14}},\ \bibinfo {pages} {103013} (\bibinfo {year} {2012})}\BibitemShut
  {NoStop}%
\bibitem [{end()}]{endnote}%
  \BibitemOpen
  \href@noop {} {}\bibinfo {note} {For Bayesian methods to work a likelihood
  for all data must be computed. Here we assume additive Gaussian noise,
  similar to what is customary to justify least-squares fitting, but choose a
  definite standard deviation. This standard deviation is chosen to be
  $\sqrt{0.1}$ which is chosen to upper bound the observed standard deviation
  over our data sets for fixed values of $N$.}\BibitemShut {Stop}%
\bibitem [{\citenamefont {Developers}(2019)}]{cirq_2019}%
  \BibitemOpen
  \bibfield  {author} {\bibinfo {author} {\bibfnamefont {The~Cirq}\
  \bibnamefont {Developers}},\ }\href {https://github.com/quantumlib/Cirq}
  {\enquote {\bibinfo {title} {Cirq},}\ } (\bibinfo {year} {2019}),\ \bibinfo
  {note} {https://github.com/quantumlib/Cirq}\BibitemShut {NoStop}%
\bibitem [{\citenamefont {Kjaergaard}\ \emph {et~al.}(2020)\citenamefont
  {Kjaergaard}, \citenamefont {Schwartz}, \citenamefont {Braum{\"u}ller},
  \citenamefont {Krantz}, \citenamefont {Wang}, \citenamefont {Gustavsson},\
  and\ \citenamefont {Oliver}}]{Kjaergaard2020-bb}%
  \BibitemOpen
  \bibfield  {author} {\bibinfo {author} {\bibfnamefont {Morten}\ \bibnamefont
  {Kjaergaard}}, \bibinfo {author} {\bibfnamefont {Mollie~E}\ \bibnamefont
  {Schwartz}}, \bibinfo {author} {\bibfnamefont {Jochen}\ \bibnamefont
  {Braum{\"u}ller}}, \bibinfo {author} {\bibfnamefont {Philip}\ \bibnamefont
  {Krantz}}, \bibinfo {author} {\bibfnamefont {Joel I-J}\ \bibnamefont {Wang}},
  \bibinfo {author} {\bibfnamefont {Simon}\ \bibnamefont {Gustavsson}}, \ and\
  \bibinfo {author} {\bibfnamefont {William~D}\ \bibnamefont {Oliver}},\
  }\bibfield  {title} {\enquote {\bibinfo {title} {Superconducting qubits:
  Current state of play},}\ }\href {\doibase
  10.1146/annurev-conmatphys-031119-050605} {\bibfield  {journal} {\bibinfo
  {journal} {Annu. Rev. Condens. Matter Phys.}\ }\textbf {\bibinfo {volume}
  {11}},\ \bibinfo {pages} {369--395} (\bibinfo {year} {2020})}\BibitemShut
  {NoStop}%
\bibitem [{\citenamefont {Bruzewicz}\ \emph {et~al.}(2019)\citenamefont
  {Bruzewicz}, \citenamefont {Chiaverini}, \citenamefont {McConnell},\ and\
  \citenamefont {Sage}}]{Bruzewicz2019-zm}%
  \BibitemOpen
  \bibfield  {author} {\bibinfo {author} {\bibfnamefont {Colin~D}\ \bibnamefont
  {Bruzewicz}}, \bibinfo {author} {\bibfnamefont {John}\ \bibnamefont
  {Chiaverini}}, \bibinfo {author} {\bibfnamefont {Robert}\ \bibnamefont
  {McConnell}}, \ and\ \bibinfo {author} {\bibfnamefont {Jeremy~M}\
  \bibnamefont {Sage}},\ }\bibfield  {title} {\enquote {\bibinfo {title}
  {Trapped-ion quantum computing: Progress and challenges},}\ }\href {\doibase
  10.1063/1.5088164} {\bibfield  {journal} {\bibinfo  {journal} {Appl. Phys.
  Rev.}\ }\textbf {\bibinfo {volume} {6}},\ \bibinfo {pages} {021314} (\bibinfo
  {year} {2019})}\BibitemShut {NoStop}%
\bibitem [{\citenamefont {Heinsoo}\ \emph {et~al.}(2018)\citenamefont
  {Heinsoo}, \citenamefont {Andersen}, \citenamefont {Remm}, \citenamefont
  {Krinner}, \citenamefont {Walter}, \citenamefont {Salath{\'e}}, \citenamefont
  {Gasparinetti}, \citenamefont {Besse}, \citenamefont {Poto{\v c}nik},
  \citenamefont {Wallraff},\ and\ \citenamefont {Eichler}}]{Heinsoo2018-tp}%
  \BibitemOpen
  \bibfield  {author} {\bibinfo {author} {\bibfnamefont {Johannes}\
  \bibnamefont {Heinsoo}}, \bibinfo {author} {\bibfnamefont
  {Christian~Kraglund}\ \bibnamefont {Andersen}}, \bibinfo {author}
  {\bibfnamefont {Ants}\ \bibnamefont {Remm}}, \bibinfo {author} {\bibfnamefont
  {Sebastian}\ \bibnamefont {Krinner}}, \bibinfo {author} {\bibfnamefont
  {Theodore}\ \bibnamefont {Walter}}, \bibinfo {author} {\bibfnamefont {Yves}\
  \bibnamefont {Salath{\'e}}}, \bibinfo {author} {\bibfnamefont {Simone}\
  \bibnamefont {Gasparinetti}}, \bibinfo {author} {\bibfnamefont {Jean-Claude}\
  \bibnamefont {Besse}}, \bibinfo {author} {\bibfnamefont {Anton}\ \bibnamefont
  {Poto{\v c}nik}}, \bibinfo {author} {\bibfnamefont {Andreas}\ \bibnamefont
  {Wallraff}}, \ and\ \bibinfo {author} {\bibfnamefont {Christopher}\
  \bibnamefont {Eichler}},\ }\bibfield  {title} {\enquote {\bibinfo {title}
  {Rapid high-fidelity multiplexed readout of superconducting qubits},}\ }\href
  {\doibase 10.1103/PhysRevApplied.10.034040} {\bibfield  {journal} {\bibinfo
  {journal} {Phys. Rev. Applied}\ }\textbf {\bibinfo {volume} {10}},\ \bibinfo
  {pages} {034040} (\bibinfo {year} {2018})}\BibitemShut {NoStop}%
\bibitem [{\citenamefont {Preskill}(2018)}]{Preskill2018-po}%
  \BibitemOpen
  \bibfield  {author} {\bibinfo {author} {\bibfnamefont {John}\ \bibnamefont
  {Preskill}},\ }\bibfield  {title} {\enquote {\bibinfo {title} {Quantum
  computing in the {NISQ} era and beyond},}\ }\href@noop {} {\bibfield
  {journal} {\bibinfo  {journal} {Quantum}\ }\textbf {\bibinfo {volume} {2}},\
  \bibinfo {pages} {79} (\bibinfo {year} {2018})}\BibitemShut {NoStop}%
\bibitem [{\citenamefont {Kitaev}(2015)}]{kitaev2015syk}%
  \BibitemOpen
  \bibfield  {author} {\bibinfo {author} {\bibfnamefont {Alexei}\ \bibnamefont
  {Kitaev}},\ }\href {http://online.kitp.ucsb.edu/online/entangled15/kitaev/}
  {\enquote {\bibinfo {title} {{A Simple Model of Quantum Holography}},}\ }
  (\bibinfo {year} {2015})\BibitemShut {NoStop}%
\bibitem [{\citenamefont {Babbush}\ \emph {et~al.}(2019)\citenamefont
  {Babbush}, \citenamefont {Berry},\ and\ \citenamefont
  {Neven}}]{babbush2019syk}%
  \BibitemOpen
  \bibfield  {author} {\bibinfo {author} {\bibfnamefont {Ryan}\ \bibnamefont
  {Babbush}}, \bibinfo {author} {\bibfnamefont {Dominic~W.}\ \bibnamefont
  {Berry}}, \ and\ \bibinfo {author} {\bibfnamefont {Hartmut}\ \bibnamefont
  {Neven}},\ }\bibfield  {title} {\enquote {\bibinfo {title} {{Quantum
  simulation of the Sachdev-Ye-Kitaev model by asymmetric qubitization}},}\
  }\href {\doibase 10.1103/PhysRevA.99.040301} {\bibfield  {journal} {\bibinfo
  {journal} {Physical Review A}\ }\textbf {\bibinfo {volume} {99}},\ \bibinfo
  {pages} {040301} (\bibinfo {year} {2019})}\BibitemShut {NoStop}%
\end{thebibliography}

\onecolumngrid
\appendix

\section{Variance Bounds}
\label{app:variance_bounds}

In the text of the main article, we reviewed the standard approach to upper
bounding the number of measurements \((M)\) required to measure the energy of a
Hamiltonian \((H)\) to within a desired target precision
\(\epsilon\)~\cite{wecker2015progress,rubin2018application}. 
This bound is calculated in a straightforward way from the qubit representation
that is obtained by making the Jordan-Wigner transformation on a second
quantized representation of the fermionic states, by expressing \(H\) as the sum
of Pauli strings (products of single-qubit Pauli operators) \(P_\ell\) acting on
the qubit representation of the state. 
Then we have
\begin{equation} M_q \leq \left(\frac{\sum_\ell
      \left|\omega_\ell\right|}{\epsilon}\right)^2, \quad \textrm{where} \quad H
  = \sum_\ell \omega_\ell P_\ell.
  \label{eq:L1_bound}
\end{equation}
More generally, if the Hamiltonian is expressed as a linear combination \(H =
\sum_\ell \omega_\ell O_\ell\), one can work out the optimal way of distributing
independent measurements between these terms and the overall number of
measurements required for the resulting estimator to attain a target precision. 
We refer the reader to Ref.~\citenum{rubin2018application} for more details and
simply recall the expression here,
\begin{equation}
  M = \left(\frac{\sum_\ell
      \left|\omega_\ell\right|\sigma_\ell}{\epsilon}\right)^2.
  \label{eq:L1_equality}
\end{equation}
The notation is the same as above, except that \(\sigma_\ell\) is the positive
square root of the variance of the operator \(O_\ell\). 
The upper bound of \eq{L1_bound} is derived by noting that the variance of a
Pauli operator measurement is at most one and by performing the appropriate
substitutions. 
Our primary concern here is to show how the calculation of such a bound for
fermionic Hamiltonians depends in a subtle manner on the representation of the
Hamiltonian. 
We have denoted the bounds above $M_q$, to refer to their evaluation in the
qubit representation of the Hamiltonian.

When calculating an upper bound of this type for a quantum chemical Hamiltonian,
\begin{equation}
  H = \sum_{pq}h_{pq} a^\dagger_p a_q + \sum_{pqrs}h_{pqrs}
  a^\dagger_p a^\dagger_q a_r a_s,
  \label{eq:qc_sq_hamiltonian}
\end{equation}
it might seem natural to work directly with the coefficients \(h_{pq}\) and
\(h_{pqrs}\) and the fermionic representation rather than performing the
Jordan-Wigner transformation to the qubit representation. 
Provided that one is careful to count the coefficient for each term and its
Hermitian conjugate only once, it is possible to obtain an upper bound to the
number of measurements required directly from the coefficients \(h_{pq}\) and
\(h_{pqrs}\). 
Provided that the Hamiltonian is normal-ordered first, this bound can be
expressed as
\begin{equation}
  M_f \leq \left( \frac{\sum_{p, q \leq p} |h_{pq}| + \sum_{pqr, s < q} |h_{pqrs}| + \sum_{pq, r \leq p, s = q} |h_{pqrs} |}{\epsilon} \right)^2.
  \label{eq:L1_bound_fermionic}
\end{equation}
We denote this bound by $M_f$ to indicate it was derived in the fermionic
representation of the Hamiltonian. 
However, this bound is looser than necessary in multiple ways. 
In \tab{norm_bound_data} we show a breakdown of the calculation of the sum of
the absolute values of the coefficients for the Hamiltonian of a chain of eight
equally spaced hydrogen atoms. 
We consider five different types of terms from \eq{qc_sq_hamiltonian} and
calculate the sum of the absolute values of the coefficients for all terms of
each type in both the qubit and fermionic representations. 
By comparing the two approaches in this way we show below that we can shed light
on the difference in the resulting bounds.

\begin{table*}[t]
\begin{tabular}{l||c|c|c|c|c|c} \hline
  Hamiltonian Partition & I & II & III & IV & V & Whole \\
                        & & & & & & Hamiltonian \\ \hline Example Term &
\(a_p^\dagger a_p\) & \(a_p^\dagger a_q + h.c.\) & \(a^\dagger_p
a^\dagger_q a_p a_q\) & \(a^\dagger_p a^\dagger_q
a_q a_r + h.c.\) & \(a^\dagger_p a^\dagger_q a_r
a_s + h.c.\) & - \\ \hline\hline Fermionic \(\sum_\ell |w_\ell|\) & 32.288
& 2.852 & 34.214 & 7.436 & 35.579 & 112.368 \\ \hline Qubit \(\sum_\ell
|w_\ell|\) & 16.144 & 2.852 & 25.660 & 6.794 & 17.790 & 33.500\\ \hline\hline
\hline
\end{tabular}
\caption{Consider the normal ordered second quantized quantum chemistry
  Hamiltonian of \eq{qc_sq_hamiltonian}, calculated for a chain of eight
  hydrogen atoms equally spaced \(1.0\)\AA~apart in an STO-3G basis. 
  We group the terms in this Hamiltonian into five partitions. 
  Partitions I and II contain the one particle terms from the first summation. 
  Partition I consists of those terms where \(p=q\), while II consists of those
  where \(p \neq q\). 
  Partitions III, IV, and V contain the two particle terms from the second
  summation. 
  Partition III consists of those where there are two unique values among p, q,
  r, and s, while IV consists of those with three unique values and V consists
  of those with four eigenvalues.
  For each partition, we report the sum of the absolute values of the
  coefficients of these terms in the fermionic representation of the Hamiltonian
  (counting the coefficient of a term and its Hermitian conjugate only once). 
  We also report the same quantity calculated in the qubit representation after
  applying the Jordan-Wigner transformation. 
  We drop any constant terms which appear as a result of the Jordan-Wigner
  transformation, since these do not contribute to the variance. 
  Additionally, we report the sum of the absolute value of the coefficients for
  the entire Hamiltonian calculated in both ways in the final column.}
  \label{tab:norm_bound_data}
\end{table*}

In column I of \tab{norm_bound_data} we begin with this analysis for the `number
operator' terms, \(a^\dagger_pa_p\) of partition I. 
We see that the value reported in the qubit representation is exactly half of
that reported in the fermionic one. 
This is because the Jordan-Wigner transformation applied to \(a^\dagger_pa_p\)
yields \(\frac{1}{2}\mathcal{I} + \frac{1}{2}Z_p\). 
However, we may neglect the first contribution since this is a constant and so
does not affect the variance. 
Another way of understanding this difference between the qubit and fermionic
contributions is to realize that the bound of \eq{L1_bound} is derived with the
assumption that the maximum variance of each term is \(1\). 
The number operator, however, has eigenvalues \(0\) and \(1\) rather than \(-1\)
and \(1\) like a Pauli operator, and so its maximum variance is lower. 
We shall present an alternative to \eq{L1_bound_fermionic} below in
\eq{L1_better_bound_fermionic} that accounts for this lower variance. 
The analysis of the other one-body term, \(a_p^\dagger a_q + h.c.\), from
partition II, is simpler. 
Column II shows that the part of the total magnitude of the coefficients of
these terms is the same, regardless of which representation is used for the
calculation.

The two-body terms in the Hamiltonian display more varied behavior. 
Taking a term from partition III and applying the Jordan-Wigner transformation
results in exactly \(\frac{1}{4}\) of the weight being assigned to a constant
term, explaining the difference between the values for this partition. 
Because terms such as \(a^\dagger_pa^\dagger_qa_pa_q\) can be rewritten as the
product of two number operators, they must have eigenvalues \(0\) and \(1\) and
thus, a maximum variance smaller than one. 
This improved bound would actually be lower than the one suggested by the
analysis of the Jordan-Wigner transformed terms, because it would account for
the covariances between the \(Z_p\), \(Z_q\), and \(Z_pZ_q\) terms that emerge. 
We shall incorporate this tighter bound on the variance of the individual terms
in this class into the alternative to \eq{L1_bound_fermionic} presented below as
\eq{L1_better_bound_fermionic}.

The disparity for partitions IV and V has a different source. 
If one performs the Jordan-Wigner transformation of a term from either class
individually, there is no difference between the total magnitudes of the
coefficients for the fermionic operators and their qubit counterparts. 
However, when one sums all such terms together there is some cancellation
between the qubit terms that reduces the overall total magnitude.
Specifically, the terms in class V benefit from some cancellation due to the
eight-fold symmetries of the two-electron integral tensor for real
orbitals~\cite{Szabo2012-ag},
\begin{equation}
h_{pqrs} = h_{qpsr} = h_{rspq} = h_{srqp} = h_{rqps} = h_{qrsp} = h_{psrq} = h_{spqr} =h.
\label{eq:h1234}
\end{equation}
We claim that the cancellations between symmetric terms result in a value for
the sum of the magnitudes of the coefficients that is exactly half as large in
the qubit representation as it is in the fermionic one. 
As an example, consider the case with \(p = 4, q = 2, r = 3, s = 1\). 
After normal-ordering, the eight terms become four and we have
\begin{equation}
 2 h a^\dagger_4 a^\dagger_2 a_3 a_1 + h.c. + 2 h a^\dagger_3 a^\dagger_2 a_4 a_1 + h.c.,
\end{equation}
where \(h\) is given by Eq.~\eqref{eq:h1234} and denotes the value of the
coefficients before normal-ordering. 
The Jordan-Wigner transformation leads to terms from \(a^\dagger_4 a^\dagger_2
a_3 a_1 + h.c.\) that cancel with terms from \(a^\dagger_3 a^\dagger_2 a_4 a_1\)
as shown below:
\begin{equation}
  \scriptsize
\begin{split}
\frac{-h}{4} \big(
X_1 X_2 X_3 X_4 +
X_1 X_2 Y_3 Y_4 -
X_1 Y_2 X_3 Y_4 +
X_1 Y_2 Y_3 X_4 +
Y_1 X_2 X_3 Y_4 -
Y_1 X_2 Y_3 X_4 +
Y_1 Y_2 X_3 X_4 +
Y_1 Y_2 Y_3 Y_4 \big) + \\  
\frac{-h}{4} \big(
X_1 X_2 X_3 X_4 +
X_1 X_2 Y_3 Y_4 +
X_1 Y_2 X_3 Y_4 -
X_1 Y_2 Y_3 X_4 -
Y_1 X_2 X_3 Y_4 +
Y_1 X_2 Y_3 X_4 +
Y_1 Y_2 X_3 X_4 +
Y_1 Y_2 Y_3 Y_4 \big) = \\
\frac{-h}{4} \big(
  X_1 X_2 X_3 X_4 +
  X_1 X_2 Y_3 Y_4 +
  Y_1 Y_2 X_3 X_4 +
  Y_1 Y_2 Y_3 Y_4 +
  X_1 X_2 X_3 X_4 +
  X_1 X_2 Y_3 Y_4 +
  Y_1 Y_2 X_3 X_4 +
  Y_1 Y_2 Y_3 Y_4 \big).
\end{split}
\end{equation}
It is straightforward, although tedious, to prove that the same cancellation
occurs generically for terms in class V as a consequence of this eight-fold
symmetry. 
As a result, for this class of terms the sum of the magnitudes of the
coefficients is exactly half as large in the qubit representation as it is in
the fermionic one. 
Analogous cancellations in the sum of the class IV terms do not show an obvious
symmetry but they are also the source of the difference between the
contributions from the two representations in partition IV.

Further cancellation is also apparent when one combines all five classes and
calculates the sum of the absolute values of the coefficients (dropping the
constant terms) for both representations of the Hamiltonian. 
The sum of magnitudes of the individual classes in the fermionic representation
is the same as the magnitude of the sum. 
However, in the qubit representation, this value calculated for the entire
Hamiltonian is roughly half the size of the sum of the individual partitions.
One substantial reason for this difference is the fact that the terms in
partitions I and III naturally give rise to terms proportional to products of
single qubit \(Z\) operators having opposite signs. 
This is behavior that we should expect for any molecular Hamiltonian, where the
single number operator terms arise from the Coulomb attraction between nuclei
and individual electrons (negative sign), while the terms containing two number
operators arise from the Coulomb repulsion between pairs of electrons (positive
sign).
Furthermore, unlike the tighter bounds achievable by accounting for the smaller
variance of the terms in partitions I and III, this cancellation derives from
the underlying form of the Hamiltonian and can not be accounted for in a
straightforward way by using a better upper bound for the \(\sigma_\ell\) values
in \eq{L1_equality} when deriving a fermionic bound like
\eq{L1_bound_fermionic}.

In the first three rows of \tab{variance_bound}, we now tabulate the bounds on
the variance of the energy estimator, in units of 100 $E_h^2$, that arise from
the sums of the absolute values of the coefficients. 
We perform these calculations for a chain of eight hydrogen atoms at various
symmetric stretched interatomic spacings, including the \(1.0\)\AA~distance
explored in \tab{norm_bound_data}. 
In this table, `Qubit Variance Bound' refers to the bound of \eq{L1_bound}
calculated using the qubit form of the Hamiltonian. 
`Naive Fermionic Variance Bound' is calculated in a similar way, except using
the sums of the fermionic coefficients, as in \eq{L1_bound_fermionic}. 
As noted above, the terms in the Fermionic Hamiltonian which consist of number
operators (class I in \tab{norm_bound_data}) or products of number operators
(class II in \tab{norm_bound_data}) actually have a variance which is
upper-bounded by \(\frac{1}{4}\) rather than one. 
One can substitute this tighter bound in \eq{L1_equality} to yield the
expression
\begin{equation}
  M_f \leq \left(\frac{\frac{1}{2} \sum_{p} |h_{p}| + \sum_{p,q < p} |h_{pq}| + \sum_{pqr, s < q} |h_{pqrs}| + \sum_{pq, r < p, q} |h_{pqrq} | + \frac{1}{2}\sum_{pq} |h_{pqpq}| }{\epsilon} \right)^2,
  \label{eq:L1_better_bound_fermionic}
\end{equation}
where we have assumed that the Hamiltonian is normal-ordered to simplify the
expression. 
We present calculation based on this improved bound in the row titled `Fermionic
Variance Bound.' 
However, it is clear that the bounds obtained directly from the qubit
representation of the Hamiltonian are considerably tighter than either of the
bounds obtained using the fermionic representation.
This difference is explained by the cancellation effects that we have described
above.

\begin{table*}[t] \footnotesize
\begin{tabular}{l||c|c|c|c|c|c|c|c} \hline Interatomic Spacing (Angstrom) & .6 & .7 & .8 &
.9 & 1.0 & 1.1 & 1.2 & 1.3 \\
  \hline\hline Qubit Variance Bound & 22.499 &
                                  17.657 &
                                           14.680 &
                                                    12.675 &
                                                             11.222 &
                                                                      10.134 &
                                                                               9.297
                                                                               &
                                                                                       8.629
                                                                                       \\
  \hline Fermionic Variance Bound &
                                    99.393 & 86.271 & 76.352 & 68.683 & 62.596 & 57.721 & 53.768 & 50.462 \\
  \hline Naive Fermionic Variance Bound & 205.251 & 177.997 & 156.704 & 139.847
   & 126.267 & 115.234 & 106.168 & 98.545 \\
  \hline \hline
Fermionic Variance Approximation & 31.501 & 27.404 & 24.580 & 22.557 & 21.038 &
19.893 & 19.023 & 18.321 \\ \hline Qubit Variance Approximation & 7.212 & 6.225
& 5.567 & 5.118 & 4.786 & 4.540 & 4.356 & 4.209 \\ \hline \hline Hartree-Fock Variance
& 7.211 & 6.224 & 5.565 & 5.118 & 4.785 & 4.539 & 4.355 & 4.208 \\ \hline Ground
State Variance & 9.206 & 8.194 & 7.568 & 7.181 & 6.929 & 6.779 & 6.695 & 6.641
\\ \hline\hline \hline
\end{tabular}
\caption{Variances for a symmetrically stretched chain of \(8\) hydrogen atoms
  in an STO-3G basis. Rows 1-5 show values of the variance bounds and approximations to these that are described in the text. The variances are presented in units of 100
  E\(_h^2\). The bound in row 1 is calculated using \eq{L1_bound}, while row 2 uses \eq{L1_better_bound_fermionic}, and row 3 uses \eq{L1_bound_fermionic}. The approximations in rows 4-5 are calculated using the methodology of
  Ref.~\citenum{wecker2015progress}, which amounts to using \eq{L1_bound} or \eq{L1_bound_fermionic} but
  neglecting some of the terms in the Hamiltonian as described in the text below. The last two rows, 6-7, present the actual variance of an
  estimator that measures each term in the Jordan-Wigner transformed
  Hamiltonian separately, for the Hartree-Fock state and for the ground state, respectively.}
\label{tab:variance_bound}
\end{table*}

In addition to these bounds on the variance of the estimator for the
Hamiltonian, we also consider two approximations to this variance that are not
guaranteed to be upper bounds (rows 4 and 5 of Table~\ref{tab:variance_bound}). 
These approximations, which we refer to for brevity as FVA and QVA, are
calculated using the methodology of Ref.~\citenum{wecker2015progress} using the
fermionic and qubit Hamiltonians respectively. 
In that work Wecker et al. 
reasoned that, in a typical quantum chemical calculation, the orbitals would
have occupation numbers near \(0\) or \(1\).
Therefore, the number and number-number terms in the Hamiltonian (partition I
and partition III in \tab{norm_bound_data}) would have a variance that is close
to zero. 
This assumption is satisfied exactly for the Hartree-Fock state when the
appropriate single-particle basis is used, and should be approximately true when
Hartree-Fock is qualitatively correct. 
Based on this assumption, Ref.~\citenum{wecker2015progress} neglected these
terms and then approximated the variance of the remaining terms in the
Hamiltonian using the type of bounds we have already discussed.

Rows 4 and 5 of Table~\ref{tab:variance_bound} show that that there is still a
substantial difference between the variances calculated under this approximation
using the two different representations of the Hamiltonian, i.e., between FVA
and QVA. 
This is primarily due to the reduction caused by the cancellations among the
double-excitation terms (class V in \tab{norm_bound_data}). 
Interestingly, the numbers presented for the `Qubit Variance Approximation'
(QVA) are nearly identical to those for the actual variance expected when
measuring the Hartree-Fock state (row 6). 
In fact, any differences between these values are found to arise purely from
numerical precision issues in the data. 
One can examine the Pauli operators that arise from performing the Jordan-Wigner
transformation on the Hamiltonian after deleting the diagonal terms and see that
each of them has an expectation value of exactly zero on the Hartree-Fock state. 
Measurements of these terms therefore achieve the maximum possible variance of
\(1\), while measurements of the deleted diagonal terms would have a variance
that is exactly \(0\). 
Thus, the calculation of the actual variance (when measuring each term in the
Hamiltonian separately) using \eq{L1_equality} for the Hartree-Fock state then
yields the same value as the calculation of the bound of \eq{L1_bound} under the
approximation proposed by Ref.~\citenum{wecker2015progress}.

\section{Applying the Fermionic RDM Constraints to the Qubit Hamiltonian}
\label{app:qubit_rdm}

In the previous section we saw a substantial difference between the bounds
calculated from the fermionic operators and those calculated from the qubit
operators after applying the Jordan-Wigner transformation. 
In light of this, it is natural to ask how the reduced density matrix (RDM)
approach of Ref.~\citenum{rubin2018application} might perform when formulated
using the qubit representation of the Hamiltonian. 
In Ref, \cite{rubin2018application}, Rubin et al. 
proposed that one could use known $n$-representability constraints on the
expectation values of few-fermion operators, in order to construct estimators
for the expectation value of the Hamiltonian that will have lower variance. 
They showed how one could take a collection of algebraic equalities from these
fermionic n-representability constraints and use them to construct a new
Hamiltonian \(\tilde{H}\) from the original \(H\). 
According to their approach, \(\tilde{H}\) is constructed to have the same
expectation value as \(H\), but a lower maximum variance according to the bounds
discussed above. 
They performed this minimization of the upper bound using standard linear
programming techniques.

We are primarily focused here on the impact of these techniques for a real-world
experiment. 
Therefore, we shall compare the impact of performing this minimization on the
fermionic and qubit representations of the Hamiltonian, using the actual
observed variance with respect to the ground state as the figure of merit,
rather than employing the bounds or approximations discussed above. 
We take the same Pauli Word Grouping strategy described in the main text and
apply it to the Hamiltonians \(\tilde{H}_{fermionic}\) and
\(\tilde{H}_{qubit}\). 
We define \(\tilde{H}_{fermionic}\) and \(\tilde{H}_{qubit}\) as the
Hamiltonians that result from performing the upper bound minimization procedure
of Ref.~\citenum{rubin2018application} in the fermionic and qubit
representations respectively.

\begin{figure*}[t]
  \includegraphics[width= .9\textwidth]{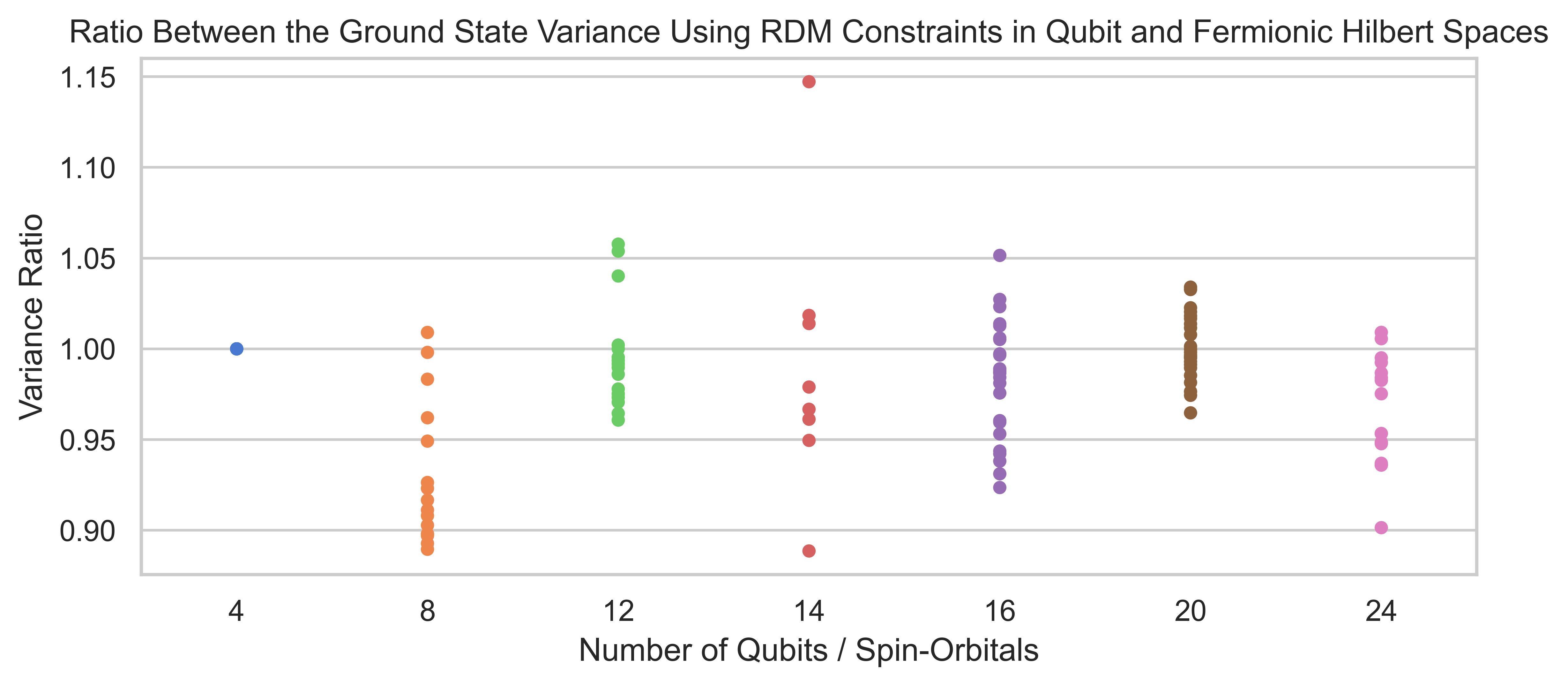}
  \caption{For each of the systems considered in the main text we apply the
    techniques of Ref.~\citenum{rubin2018application} to the Hamiltonians in the
    fermionic and qubit Hibert spaces. 
    We list these systems in \tab{system_list} below, duplicating Table II of
    the main text for convenience. 
    Using fermionic n-representability constraints, we construct the
    Hamiltonians \(\tilde{H}_{fermionic}\) and \(\tilde{H}_{qubit}\), that have
    the same expectation value but a lower maximum variance under bounds of the
    type described by \eq{L1_bound} and \eq{L1_bound_fermionic}. 
    We then consider the variance of these Hamiltonians with respect to the
    ground state. 
    We calculate these variances assuming measurement is performed using the
    Pauli Word Grouping strategy described in the main text. 
    Finally, we plot the ratio of the variance obtained for
    \(\tilde{H}_{qubit}\) with the variance obtained for
    \(\tilde{H}_{fermionic}\). 
    The fact that all of these ratios are found to be near \(1\) shows that
    reformulating the work of Ref.~\citenum{rubin2018application} in the qubit
    representation does not offer a substantial improvement.
  }
  \label{fig:rdm_ratio}
\end{figure*}

In \fig{rdm_ratio} we plot the ratio between the variances of
\(\tilde{H}_{qubit}\) and \(\tilde{H}_{fermionic}\) for the ground state of each
of the systems considered in the main text of this work. 
We list these systems in \tab{system_list} below, duplicating Table II of the
main text for convenience. 
Despite the substantial differences in the variance bounds formulated in the two
representations, the impact of applying the RDM constraints to the qubit
Hamiltonian rather than the fermionic one is found to be marginal, at best. 
For the majority of the systems it appears that the qubit-based bounds perform
slightly better, but there are also a number of cases where this pattern is
reversed.

\section{Low Rank Decomposition}
\label{app:low_rank_decomp}

In the main text, we explained that our strategy for measurement is based on
rewriting the standard quantum chemical Hamiltonian in the following form:
\begin{equation}
  H = U_0 \left(\sum_{p} g_{p} n_{p} \right) U_0^\dagger + \sum_{\ell=1}^L
  U_\ell \left(\sum_{p q} g_{p q}^{(\ell)} n_{p} n_{q}\right) U_\ell^\dagger,
  \label{eq:decomposed_ham}
\end{equation}
where the values $g_p$ and $g_{pq}^{(\ell)}$ are scalars, $n_p = a^\dagger_p
a_p$, and the \(U_\ell\) are unitary operators which implement a single particle
change of orbital basis. 
Here we shall explain how one obtains that factorization starting from a
standard representation. 
We follow the presentation of Berry et al. 
with minor deviations and refer the reader to Ref.~\citenum{Berry2019-ux} for
more details.

First, it is necessary to obtain the Hamiltonian in the chemist's
standard form,
\begin{equation}
  H = \sum_{\sigma \in \{\uparrow, \downarrow\}}\sum_{pq}T_{pq} a^\dagger_{p,\sigma} a_{q, \sigma} + \frac{1}{2} \sum_{\alpha, \beta \in \{\uparrow, \downarrow\}} \sum_{pqrs}V_{pqrs}
  a^\dagger_{p,\alpha} a_{q, \alpha} a^\dagger_{r, \beta} a_{s, \beta}.
  \label{eq:chemist_hamiltonian}
\end{equation}
This differs from the physicist's convention of \eq{qc_sq_hamiltonian}, where
the operators in the two-electron component of the Hamiltonian have both
creation operators to the left of both annihilation operators. 
We assume the use of purely real spatial orbitals, and therefore the tensor
\(V_{pqrs}\) inherits the eight-fold symmetry,
\begin{equation}
V_{pqrs} = V_{srqp} = V_{psqr} = V_{qprs} = V_{qpsr} = V_{rsqp} = V_{rspq} = V_{srpq},
\end{equation}
from the definition of the two-electron integrals~\cite{Szabo2012-ag}.

Now we can perform the decomposition. We treat the tensor \(V\) as a matrix indexed by the
collective indices \(pq\) and \(rs\). We can eigendecompose this matrix to
yield
\begin{equation}
  V_{pqrs} = \sum_\ell w_\ell v^{(\ell)}_{pq} v^{(\ell)}_{rs}
\end{equation}
In the above equation, \(w_{\ell}\) are the eigenvalues of \(V\), \(v^{(\ell)}\)
are the eigenvalues. We proceed by using this equality to rewrite the two-electron component of the
Hamiltonian,
\begin{equation}
  \frac{1}{2} \sum_{\alpha, \beta \in \{\uparrow, \downarrow\}} \sum_{pqrs}V_{pqrs}
  a^\dagger_{p,\alpha} a_{q, \alpha} a^\dagger_{r, \beta} a_{s, \beta} =
  \frac{1}{2} \sum_\ell w_\ell\Big( \sum_{\sigma \in \{\uparrow, \downarrow\}} \sum_{pq} v^{(\ell)}_{pq} a^\dagger_{p, \sigma} a_{q, \sigma} \Big)^2,
  \label{eq:mostly_decomposed}
\end{equation}
with \(v^{(\ell)}_{pq}\) inheriting the symmetry between the \(p\) and \(q\)
indices from \(V\).

The final remaining step is to transform \eq{mostly_decomposed}, as well as the
one-electron component of the Hamiltonian by diagonalizing each of the one-body
operators. 
It is straightforward to express each of the one-body operators as diagonal
operators in a rotated single-particle basis. 
The appropriate change of basis matrices can be obtained from the eigenvalues of
the coefficient tensors, \(T\) and the \(g^{(\ell)}\)s in our case. 
We can therefore express the Hamiltonian in the form of \eq{decomposed_ham},
dropping the spin indices for simplicity. 
The coefficients \(g_p\) come from rotating to a basis where \(T_{pq}\) is
diagonal. 
The coefficients \(g^{(\ell)}_{pq}\) likewise come from rotating to a series of
bases where the tensors \(v^{\ell}\) are diagonal between their \(p\) and \(q\)
indices. 
The operators \(U_\ell\) are the inverse of the operators that diagonalize the
one-body operators \(\sum_{\sigma \in \{\uparrow, \downarrow\}}\sum_{pq}T_{pq}
a^\dagger_{p,\sigma} a_{q, \sigma}\) and \( \sum_{\sigma \in \{\uparrow,
  \downarrow\}} \sum_{pq} v^{(\ell)}_{pq} a^\dagger_{p, \sigma} a_{q, \sigma}
\). 
Note that the \(p\) and \(q\) indices of \eq{decomposed_ham} represent new dummy
indices and that the \(w_\ell\) terms have been absorbed into
\(g^{(\ell)}_{pq}\), together with the contributions from the squares of the
diagonalized \(v^{\ell}_{pq}\) terms.

\section{Description of Data}
\label{app:data_desc}
In addition to this appendix, we also include the raw data we have
generated through our numerical calculations which does not already appear in
tables throughout the manuscript. 
The data is provided as a table in the csv file format with each row
corresponding to one of the particular systems listed below in \tab{system_list}
(identical to Table II in the main text) and each column corresponding to the
variance of a different estimator (or an ancillary piece of data, such as the
energy of the system). 
Energies are provided in units of \(E_h\) and variances in units of \(E_h^2\).

\end{document}